\documentclass[aps,showpacs,twocolumn,twoside,groupedaddress]{revtex4}
\usepackage{graphicx}
\usepackage{epsfig}
\usepackage{epsf}
\usepackage{amssymb}
\usepackage{amsmath}
\usepackage[toc]{appendix}
\usepackage{dcolumn}
\usepackage{bm}

\begin{document}
	

\title{Multiphoton pulses interacting with multiple emitters in a one-dimensional waveguide}

\author{Zeyang Liao$^{1}$\footnote{ liaozy7@mail.sysu.edu.cn}, Yunning Lu$^{1}$,  and M. Suhail Zubairy$^{2}$}
\affiliation{$^1$ School of Physics, Sun Yat-sen University, Guangzhou 510275, the People's Republic of China   \\
	$^{2}$ Institute for Quantum Science and Engineering (IQSE) and Department of Physics and Astronomy, Texas A$\&$M University, College Station, TX 77843-4242, USA }

\begin{abstract}
We derive a  generalized master equation for multiphoton pulses interacting with multiple emitters in a waveguide-quantum electrodynamics system where the emitter frequency can be modulated and the effects of non-guided modes can also be considered. Based on this theory, we can calculate the real-time dynamics of an array of interacting emitters driven by an incident photon pulse which can be vacuum, a coherent state,  a Fock state or their superpositions. Moreover, we also derive generalized input-output relations to calculate the reflectivity and transmissivity of this system. We can also calculate the output photon pulse shapes.  Our theory can find important applications in the study of waveguide-based quantum systems.
\end{abstract}

 \maketitle

\section{Introduction}

In recent years, there has been a great interest in building a large scale quantum network from single quantum system \cite{Kimble2008}. However it remains a big challenge to manufacture a single quantum system which contains a large number of qubits. In contrast, it is relatively easy to first build a small quantum system with high precision controls and interconnect these small systems into a large scale quantum network via certain quantum channels. Waveguide quantum electrodynamics (QED), which studies the interaction between emitters and waveguide photons, is a good system for realizing large scale quantum network. A large number of studies have been made in the past two decades in the field of waveguide QED \cite{Zheng2013, Liao-review, Roy-review} . The emitter-photon interaction can be significantly enhanced in the reduced dimensions and the emission of an emitter can be collected by the waveguide with nearly unit efficiency \cite{Arcari2014}. This can have important applications in producing highly efficient single photon sources \cite{Laucht2012}, single photon detection \cite{Schuck2013}, and formation of atom cavity \cite{Zhou2008, Dong2009, Chang2012, Liao2016a}. The emitter-emitter interaction mediated by the one-dimensional (1D) waveguide modes can be long-range. This provides a unique system for studying many-body physics \cite{Douglas2015, Richerme2014}  and long-range quantum information transfer \cite{Yu2020, Leent2020}. Due to the confinement of transverse field, the photon modes in a quasi-1D waveguide can also have intrinsic direction dependent longitudinal angular momentum \cite{Bliokh2015}  which is suitable for studying chiral quantum optics \cite{Pichler2015, Lodahl2017, Petersen2014, Mahmoodian2016, Cheng2016, Xia2018,Lu2018}. The waveguide-QED theory can be applied to a number of systems such as the photonic line defects coupling to quantum dots \cite{Grim2019}, cold atoms trapped along the alligator waveguide \cite{Douglas2016}, superconducting qubits interacting with the microwave transmission lines \cite{Arute2019, Song2019}, and the plasmonic nanowire coupling to quantum emitters \cite{Tame2013}.

The theory of single photon transport provides the basics for studying the waveguide-QED system. Shen and Fan used the real-space Hamiltonian together with the Bethe-ansatz to study the stationary properties of single photon scattering by a single quantum emitter coupled to a 1D waveguide \cite{Shen2005} . This method was then extended to multi-emitter \cite{Tsoi2008, Cheng2017} and multi-level systems where many interesting effects can occur such as photon frequency conversion \cite{Bradford2012, Wang2014,Lu2017}, electromagnetic induced transparency (EIT) \cite{Roy2011}, realization of single photon transistor \cite{Witthaut2010} and single photon switch \cite{Shomroni2014}. In addition to the stationary spectrum and the real-time dynamics of the emitter system are also very interesting as the emitters are important units for quantum information processing and storage. Chen et al. applied the wavefunction approach to study the dynamics of a single photon pulse interacting with a single emitter \cite{Chen2011}. We generalized the wavefunction approach to the situations of multiple identical \cite{Liao2015} and non-identical
\cite{Liao2016} emitters. This allowed us to study many interesting applications in the collective many-body physics and quantum information such as quantum state preparation \cite{Liao2018}  and
waveguide-based quantum sensing \cite{Liao2017}. Recently, Dinc et al. developed an analytical method based on Bethe-ansatz approach to study the time dynamics of a single photon transport problem \cite{Dinc2019}.

Compared to the single photon problem, the multiphoton transport problem can provide more interesting physics. However calculation becomes much more complicated. The Bethe-ansatz approach can be extended to calculate the few-photon scattering problem where photon-photon bound states can occur \cite{Shen2007, Shi2016a, Facchi2016,Calajo2019}. However, when this method is generalized to more than two photons, the calculation becomes extremely cumbersome \cite{Zheng2010, Fang2014, Shen2015, Fang2017}. Alternative methods such as Lehmann-Symanzik-Zimmermann reduction method \cite{Shi2009, Shi2011}, the Green function decomposition of multiple particle scattering matrix \cite{Laakso2014}, the input-output formalism \cite{Fan2010, Lalumiere2013, Xu2015, Combes2017}, the Feynman diagrams \cite{Roulet2016, Dinc2020}, and the SLH formalism \cite{Brod2016, Combes2018} have also been proposed. In these stationary state calculations, the photons are usually assumed to be a plane wave and the real-time dynamics of the emitters are usually ignored. Based on Heisenberg-Langevin approach, Domokos et al. studied the coherent photon pulse scattering by a single quantum emitter
 in a 1D waveguide \cite{Domokos2002}. Chumak and Stolyarov investigated the propagation of few-photon pulses interacting with a single two-level system by the method of distribution functions in coordinate-momentum space \cite{Chumak2013, Chumak2014}.  Kony and Gea-Banacloche generalized the wavefunction approach to study the one- and two-photon scattering by two emitters coupled to a
1D waveguide \cite{Konyk2017}. In 2015, Caneva et al. used the effective Hamiltonian approach to derive a master equation to calculate the emitter dynamics driven by a coherent photon pulse \cite{Caneva2015}, and this method can be generalized to various systems \cite{Shi2015, Song2018}. In Refs. \cite{Shi2015, Song2018}, although they mentioned that the Fock states can be expressed as derivations of coherent state, but it usually requires that the system dynamics has analytical solutions. Otherwise, it is very difficult to directly calculate the dynamics with the Fock state input. For the continuous-mode Fock state inputs, Gheri et al. derived a mater equation to study the dynamics of a single emitter driven by a single and two photon wavepackets \cite{Gheri1998}. Baragiola et. al generalized this method to the general N-photon case based on the Ito-Langevin approach \cite{Baragiola2012} where they mainly focused on the scattering of a single emitter system. By modeling the input pulse as the output of a virtual cavity with time-dependent coupling strength, the probability to generate a specific prechosen output field can be calculated from the master equation of a virtual cascaded system \cite{Kiilerich2019}.   In 2018, we derived a master equation to study the dynamics of multiple emitters driven by continuous squeezed vacuum field in 1D waveguide \cite{You2018}  and found that steady-state population inversion of multiple $\Xi$-type emitters can occur in this system \cite{You2019}.

In this article, we explicitly derive a generalized master equation to study the dynamics of multiphoton pulses in a 1D waveguide interacting with multiple emitters whose transition frequencies can be modulated and the effects of non-guided modes are also considered. Moreover, we also derive a generalized input-output theory to calculate the output photon pulse shapes and the reflectivity/transmissivity of this system for various input photon fields. In particular, our theory allows to study the collective dynamics of multi-emitter systems driven by non-classical photon pulses which is seldom studied before. The theory developed here can thus find important applications in the researches of waveguide-based quantum system.

This article is arranged as follows. In Sec. II, we derive a generalized master equation for the emitter dynamics and present a generalized input-output theory to study the scattering field properties. In Sec. III, we apply this theory to the cases of coherent state, single and general N photon state inputs. Finally, we summarize our results.

\section{multiphoton scattering theory}

In this section, we first derive a generalized master equation for general multiphoton pulses interacting with multiple emitters coupled to a 1D waveguide. Then we derive a generalized input-output theory to calculate the reflection and transmission properties of the photon field. 

\subsection{Generalized master equation for arbitrary photon input}

\begin{figure}
	\includegraphics[width=0.9\columnwidth]{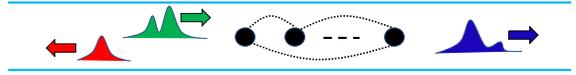}
	\caption{Multiphoton wavepacket interacting with multiple emitters in a 1D waveguide.}
	\label{1}
\end{figure}

The model we study in this paper is shown in Fig. 1. A light pulse which may contain multiple photons is injected into a 1D waveguide and it can interact with $N_{a}$ emitters with arbitrary spatial distributions. The emitter positions are denoted as $\vec{r}_{j}$ and their z-components are  $z_{j}$ where $j=1,2,\cdots, N_a$. Here, we consider a general case where the emitters can have time-modulated transition frequencies and they can couple to both the waveguide and non-waveguide photon modes. It is convenient  to work in the rotating frame with the original emitter  frequency $\omega_{a}$.
The total Hamiltonian of the system and reservoir fields in the rotating frame is given by 
\begin{align}
H(t)=&\frac{\hbar}{2}\sum_{j=1}^{N_{a}}\varepsilon_{j}(t)\sigma_{j}^{z} +\hbar\sum_{k}\Delta\omega_{k}a_{k}^{\dagger}a_{k}+\hbar\sum_{\vec{q}_{\lambda} }\Delta\omega_{\vec{q}_{\lambda}}a_{\vec{q}_{\lambda}}^{\dagger}a_{\vec{q}_{\lambda}} \nonumber\\
&+\hbar\sum_{j=1}^{N_a}\sum_{k}(g_{k}^{j}e^{ikz_{j}}\sigma_{j}^{+}a_{k}+H.c.) \nonumber\\
&+ \hbar\sum_{j=1}^{N_a}\sum_{\vec{q}_{\lambda} }(g_{\vec{q}_{\lambda} }^{j}e^{i\vec{q}\cdot \vec{r}_{j}}\sigma_{j}^{+}a_{\vec{q}_{\lambda}}+H.c.).
\end{align}
The physical meaning of each term in the Hamiltonian is as follows. The first term is the emitter Hamiltonian with time-dependent modulating frequency $\varepsilon_{j}(t)$. $\sigma_{j}^{z}$ and $\sigma_{j}^{+}$ ($\sigma_{j}^{-}$) are the zth component and the raising (lowering) Pauli operators of the $j$th emitter. The second term is the Hamiltonian of the waveguide photons with the detuning frequency $\Delta\omega_{k}=\omega_{k}-\omega_{a}$ and $a_{k} (a_{k}^{\dagger})$ is the annihilation (creation) operator of the waveguide photon mode with frequency $\omega_{k}$. When the emitter transition frequency is not very close to the photonic band edge, linear dispersion approximation can be applied where  $\Delta\omega_{k}=\delta k v_{g}$ with $\delta k=k-k_{a}$ and $v_{g}$ is the group velocity of the guided photon pulse.   The third term is the non-guided reservoir field Hamiltonian  where $a_{\vec{q}_{\lambda}} (a_{\vec{q}_{\lambda}}^{\dagger})$ is the annihilation (creation) operator of the non-guided photon mode with frequency $\omega_{\vec{q}_{\lambda}}$ ($\vec{q}$ is the wavevector and $\lambda$ denotes the polarization) and $\Delta\omega_{\vec{q}_{\lambda}}=\omega_{\vec{q}_{\lambda}}-\omega_{a}$. The fourth term is the emitter-waveguide photon interaction Hamiltonian with $g_{k}^{j}=\vec{\mu}_{j}\cdot \vec{E}_{k}(\vec{r}_{j})/\hbar$ being the coupling strength where $\vec{\mu}_{j}$ is the transition dipole moment of the jth emitter, and $\vec{E}_{k}(\vec{r}_{j})$ is the strength of the electric field with wavevector $k$ at position $\vec{r}_{j}$. The last term describes the interaction between the emitters and the non-guided reservoir field with coupling strength $g_{\vec{q}_{\lambda}}^{j}=\vec{\mu}_{j}\cdot \vec{E}_{\vec{q}_{\lambda}}(\vec{r}_{j})/\hbar$. $\hbar $ is the Planck constant. We should note that rotating wave approximation is applied in the Hamiltonian shown in Eq. (1). Thus, our theory develop here is valid for the weak and strong coupling regime, while it is invalid  in the ultrastrong coupling regime where the rotating wave approximation may break down.

According to the Heisenberg equation, the dynamics of an arbitrary emitter operator $O_{s}$ is given by
\begin{align}
\dot{O}_S(t)=&\frac{i}{2}\sum_{j=1}^{N_{a}}\varepsilon_{j}(t)[\sigma_{j}^{z}(t),O_{S}(t)] \nonumber \\
&+  i\sum_{j=1}^{N_a}\sum_{k}\Big(g_{k}^{j}e^{ikz_{j}}[\sigma_{j}^{+}(t),O_{S}(t)]a_{k} \nonumber \\
&+g_{k}^{j*}e^{-ikz_{j}}a_{k}^{\dagger}[\sigma_{j}^{-}(t),O_{S}(t)]\Big) \nonumber \\
&+i\sum_{j=1}^{N_a}\sum_{\vec{q}_{\lambda}}\Big(g_{\vec{q}_{\lambda}}^{j}e^{i\vec{q}\cdot \vec{r}_{j}}[\sigma_{j}^{+}(t),O_{S}(t)]a_{\vec{q}_{\lambda}}\nonumber \\
&+g_{\vec{q}_{\lambda}}^{j*}e^{-i\vec{q}\cdot \vec{r}_{j}}a_{\vec{q}_{\lambda}}^{\dagger}[\sigma_{j}^{-}(t),O_{S}(t)]\Big), 
\end{align}
and the dynamics of the field operators are given by
\begin{align}
\dot{a}_k(t)&=-i\Delta\omega_{k}a_{k}-i\sum_{j=1}^{N_a} g_{k}^{j*}e^{-ikz_{j}}\sigma_{j}^{-}(t), \\
\dot{a}^{\dagger}_k(t)&=i\Delta\omega_{k}+i\sum_{j=1}^{N_a} g_{k}^{j}e^{ikz_{j}}\sigma_{j}^{+}(t),\\
\dot{a}_{\vec{q}_{\lambda}}(t)&=-i\Delta\omega_{\vec{q}_{\lambda}}a_{\vec{q}_{\lambda}}-i\sum_{j=1}^{N_a} g_{\vec{q}_{\lambda}}^{j*}e^{-i\vec{q}\cdot \vec{r}_{j}}\sigma_{j}^{-}(t), \\
\dot{a}^{\dagger}_{\vec{q}_{\lambda}}(t)&
=i\Delta\omega_{\vec{q}_{\lambda}}a_{\vec{q}_{\lambda}}^{\dagger}+i\sum_{j=1}^{N_a} g_{\vec{q}_{\lambda}}^{j}e^{i\vec{q}\cdot \vec{r}_{j}}\sigma_{j}^{+}(t).
\end{align}
Formally integrating Eqs. (3-6), we can obtain
\begin{align}
a_{k}(t)=&a_{k}(0)e^{-i\Delta\omega_{k}t} \nonumber\\
&-i\sum_{j=1}^{N_a}g_{k}^{j*}e^{-ikz_{j}}\int_{0}^{t}\sigma_{j}^{-}(t')e^{i\Delta\omega_{k}(t'-t)}dt', \\
a_{k}^{\dagger}(t)=&a_{k}^{\dagger}(0)e^{i\Delta\omega_{k}t} \nonumber\\
&+i\sum_{j=1}^{N_a}g_{k}^{j}e^{ikz_{j}}\int_{0}^{t}\sigma_{j}^{+}(t')e^{-i\Delta\omega_{k}(t'-t)}dt',\\
a_{\vec{q}_{\lambda}}(t)=&a_{\vec{q}_{\lambda}}(0)e^{-i\Delta\omega_{\vec{q}_{\lambda}}t} \nonumber\\
&-i\sum_{j=1}^{N_a}g_{\vec{q}_{\lambda}}^{j*}e^{-i\vec{q}\cdot \vec{r}_{j}}\int_{0}^{t}\sigma_{j}^{-}(t')e^{i\Delta\omega_{\vec{q}_{\lambda}}(t'-t)}dt', \\
a_{\vec{q}_{\lambda}}^{\dagger}(t)=&a_{\vec{q}_{\lambda}}^{\dagger}(0)e^{i\Delta\omega_{\vec{q}_{\lambda}}t} \nonumber\\
&+i\sum_{j=1}^{N_a}g_{\vec{q}_{\lambda}}^{j}e^{i\vec{q}\cdot \vec{r}_{j}}\int_{0}^{t}\sigma_{j}^{+}(t')e^{-i\Delta\omega_{\vec{q}_{\lambda}}(t'-t)}dt',
\end{align}
from which we can see that the field at time $t$ is the interference between the incident field and the emitted field by the emitters.  
Inserting Eqs. (7-10) into Eq. (2) and using the Weisskopf-Wigner approximation we can obtain (see Appendix A)
\begin{align}
\dot{O}_S(t)=&\frac{i}{2}\sum_{j=1}^{N_{a}}\varepsilon_{j}[\sigma_{j}^{z}(t),O_{S}(t)] \nonumber\\
&+ i\sum_{j=1}^{N_a} \sqrt{\frac{\Gamma_{j}}{2}}[\sigma_{j}^{+}(t),O_{S}(t)][a_{j}(t)+b_{j}(t)] \nonumber\\
&+i\sum_{j=1}^{N_a}\sqrt{\frac{\Gamma_{j}}{2}}[a_{j}^{\dagger}(t)+b_{j}^{\dagger}(t)][\sigma_{j}^{-}(t),O_{S}(t)]  \nonumber \\ 
&+\sum_{jl}\Lambda_{jl}[\sigma_{j}^{+}(t),O_{S}(t)]\sigma_{l}^{-}(t) \nonumber\\
&-\sum_{jl}\Lambda_{jl}^{*}\sigma_{l}^{+}(t)[\sigma_{j}^{-}(t),O_{S}(t)],
\end{align}
where $a_{j}(t)=\sqrt{\frac{v_{g}}{2\pi}}\int_{-\infty}^{\infty}e^{ikz_{j}}a_{k}(0)e^{-i\delta\omega_{k}t}dk$ 
describes the absorption of the incident waveguide photons and $b_{j}(t)=\sqrt{\frac{v_{g}}{2\pi}}\int \int\int e^{i\vec{q}_{\lambda}\cdot \vec{r}_{j}}a_{\vec{q}_{\lambda}}(0)e^{-i\delta\omega_{\vec{q}_{\lambda}}t}d^{3}\vec{q}_{\lambda}$ is the absorption of the incident nonguided photons. The collective interaction between the emitters can be calculated as \cite{Liao2016}
\begin{align}
\Lambda_{jl}=&\frac{\sqrt{\Gamma_{j}\Gamma_{l}}}{2}e^{ik_{a}|z_{jl}|} +\frac{3\sqrt{\gamma_{j}\gamma_{l}}}{4}\Big[\sin^{2}\phi\frac{-i}{k_{a}r_{jl}} \nonumber\\
&+(1-3\cos^{2}\phi)(\frac{1}{(k_{a}r_{jl})^2}+\frac{i}{(k_{a}r_{jl})^{2}})\Big]e^{ik_{a}|r_{jl}|},
\end{align}
where the first term is the effective interaction mediated by the waveguide photons and the second term is the usual dipole-dipole interaction  induced by the non-guided reservoir fields.  $|r_{jl}|=|\vec{r_{j}}-\vec{r_{l}}|$ is the distance between the jth and lth emitters and $|z_{jl}|=|\vec{z_{j}}-\vec{z_{l}}|$ is the distance in the zth direction.
$\Gamma_{j}=4\pi |g_{k_{a}}^{j}|^{2}/v_{g}$ is the decay rate due to the waveguide vacuum field  and $\gamma_{j}$ is the spontaneous decay rate due to the nonguided  photon modes.  $\phi$ is the angle between the direction of the transition dipole moment and the waveguide direction. In deriving Eq. (11), we have neglected the time-retarded effects which is a good approximation when the largest emitter separation is not very large. To be more specific, the photon propagation time through the system should be much less than the decay time of the emitter (i.e., $\text{Max}(z_{ij})/v_{g}\ll 1/\Gamma_{i,j}$). Indeed, this is the usual case. For example, if $v_g\sim 10^{8}m/s$ and $\Gamma\sim 10^{8}Hz$, we require that the largest distance between the emitters is much less than $1m$ which is the usual case. 

From Eq. (11), we can derive a corresponding master equation for the emitters. Since $Tr_{S+R}[O_S(t)\rho]=Tr_S[O_S\rho_S(t)]$ where $\rho_S(t)=Tr_{R}[\rho(t)]$ is the emitter system density operator,  by time derivation on both sides we have $Tr_S[O_S\dot{\rho_S}(t)]=Tr_{S+R}[\dot{O}_S(t)\rho]$ and from Eq. (11)  we can obtain (see Appendix A)
\begin{widetext}
\begin{align}
\dot{\rho}_S(t)=& -\frac{i}{2}\sum_{j=1}^{N_{a}}\varepsilon_{j}(t)[\sigma_{j}^{z},\rho_{S}(t)]-i\sum_{j=1}^{N_a}\sqrt{\frac{\Gamma_{j}}{2}} [\sigma_{j}^{+},\rho'_{j}(t)]  -i\sum_{j=1}^{N_a}\sqrt{\frac{\Gamma_{j}}{2}}[\sigma_{j}^{-},{\rho'}_{j}^{\dagger}(t)] -i\sum_{jl}\text{Im}(\Lambda_{jl})[\sigma_j^{+}\sigma_l^{-},\rho_{S}(t)] \nonumber\\
&-\sum_{jl}\text{Re}(\Lambda_{jl})[\sigma_{j}^{+}\sigma_{l}^{-}\rho_{S}(t)
+\rho_{S}(t)\sigma_{j}^{+}\sigma_{l}^{-}-2\sigma_{l}^{-}\rho_{S}(t)\sigma_{j}^{+}].  
\end{align}
\end{widetext}
where $\rho'_{j}(t)=Tr_{R}\{U(t)[a_{j}(t)+b_{j}(t)]\rho(0) U^{\dagger}(t)\}$ is a new operator appearing in the equation.
Here, we consider the case that the incident phontons are from the waveguide photons while the nonguided reservoir field is initially in the vacuum. Since $a_{\vec{q}_{\lambda}}|0\rangle=0$, we have $b_{j}(t)\rho(0)=0$. Then we have $\rho'_{j}(t)=Tr_{R}[U(t)a_{j}(t)\rho(0) U^{\dagger}(t)]$ which accounts for the driving of the incident waveguide photons. This is the main equation of this section.  

The first term in Eq. (13) describes the modulation of the emitter transition frequencies.  The second and third terms describe the excitation and deexcitation due to the incident photon field. 
The forth term describes the  dipole-dipole interactions between the emitters induced by the guided and nonguided vacuum field. The last term is the collective dissipation due to the guided and nonguided vacuum fluctuation. However, we should note that Eq. (13) itself is in general not closed because we have the new operators like $\rho'_{j}(t)$ and  ${\rho'}_{j}^{\dagger}(t)$. In some special cases, Eq. (13) is closed. For example, if there is not external driving field, the second and third terms disappear and the equation is closed from which the emitter excitation transport can be studied. Another example is that if the incident field is a coherent field or superposition of coherent fields, the $\rho'_{j}(t)$ and ${\rho'}_{j}^{\dagger}(t)$ terms can then be reduced to a complex number multiplying $\rho_{S}(t)$ and Eq. (13) becomes closed again from which  the full dynamics of the emitters driven by a coherent field can be calculated. For the coherent state input, the master equation shown in Eq. (13) is reduced to the results shown in Ref. \cite{Caneva2015} when there is not frequency modulation and the effects of non-guided modes are ignored. 
In general cases such as the Fock state input, we have to repeat the above procedures to derive equations for $\rho'_{j}(t)$ until all the equations are closed.  For the Fock state input,  Eq. (13) is reduced to the results shown in Ref. \cite{Gheri1998} when there is only a single emitter.

\subsection{The generalized input-output theory}

In the previous subsections, we derive the master equations for the emitter system which allows to calculate the real dynamics of the emitters for an arbitrary photon wavepacket input.  In this subsection, we derive the generalized input-output relations of this system by expressing the output field operators as the function of input operators and the system operators. Together with the master equations derived in the previous subsection, we can then study the  reflection and transmission properties of this system.

If we integrate Eq. (3) from $t$ to $t_{f}$ where $t_f>t$, we can obtain
\begin{align}
a_{k}(t)=&a_{k}(t_f)e^{i\Delta\omega_{k}(t_f-t)} \nonumber\\
&+i\sum_{j=1}^{N_a}g_{k}^{j*}e^{-ikz_{j}}\int_{t}^{t_f}\sigma_{j}^{-}(t')e^{i\Delta\omega_{k}(t'-t)}dt'.
\end{align}
Comparing Eq. (7) with Eq. (14) it is readily to obtain that
\begin{align}
&a_{k}(t_f)e^{i\Delta\omega_{k}(t_f-t)} \nonumber\\
=&a_{k}(0)e^{-i\Delta\omega_{k}t} -i\sum_{j=1}^{N_a}g_{k}^{j}e^{-ikz_{j}}
\int_{0}^{t_f}\sigma_{j}^{+}(t')e^{i\Delta\omega_{k}(t'-t)}dt'.
\end{align}

We can define the following input-output operators \cite{Walls2008}
\begin{align}
a_{in}^{R}(t)&=\sqrt{\frac{v_{g}}{2\pi}}\int_{0}^{\infty}a_{k}(0)e^{-i\Delta\omega_{k}t}dk, \\
a_{in}^{L}(t)&=\sqrt{\frac{v_{g}}{2\pi}}\int_{-\infty}^{0}a_{k}(0)e^{-i\Delta\omega_{k}t}dk,  \\
a_{out}^{R}(t)&=\sqrt{\frac{v_{g}}{2\pi}}\int_{0}^{\infty}a_{k}(t_f)e^{i\delta kz_{N}}e^{-i\Delta\omega_{k}(t-t_f)}dk, \\
a_{out}^{L}(t)&=\sqrt{\frac{v_{g}}{2\pi}}\int_{-\infty}^{0}a_{k}(t_f)e^{-i\delta kz_{1}}e^{-i\Delta\omega_{k}(t-t_f)}dk,
\end{align}
where $z_{1}$ is the position of the left most emitter and $z_{N}$ is the position of the right most emitter. Since the right output field propagates freely after scattering by th right most emitter and the left output field propagates freely after scattering by the first emitter, phase factors $e^{i\delta kz_{N}}$ and $e^{-i\delta kz_{1}}$ are added in the definitions of the right and left output operators, respectively \cite{Caneva2015}.  
From Eq. (15) we can obtain the generalized input-output relations (see Appendix B)
\begin{align}
a_{out}^{R}(t)&= a_{in}^{R}(t-z_{N}/v_{g})-i\sum_{j=1}^{N_a}\sqrt{\frac{\Gamma_{j}}{2}}e^{-i k_{a}z_{j}}\sigma_{j}^{-}(t),\\
a_{out}^{L}(t)&=a_{in}^{L}(t+z_{1}/v_{g})-i\sum_{j=1}^{N_a}\sqrt{\frac{\Gamma_{j}}{2}}e^{i k_{a}z_{j}}\sigma_{j}^{-}(t), 
\end{align}
where $z_{Nj}=z_{N}-z_{j}$. From these two generalized input-output relations we can calculate the properties of the scattering field  of this system. We can define the instant field intensity propagating to the right and to the left at time $t$ by $r(t)=\langle a_{out}^{R+}(t)a_{out}^{R}(t)\rangle$ and $l(t)=\langle a_{out}^{L+}(t)a_{out}^{L}(t)\rangle $, respectively, which are given by 
\begin{align}
r(t)=& \langle a_{in}^{R+}(t-z_{N}/v_{g})a_{in}^{R}(t-z_{N}/v_{g})\rangle \nonumber\\
&-2 \sum_{j=1}^{N_a}\sqrt{\frac{\Gamma_{j}}{2}}\text{Im}[e^{i k_{a}z_{j}}\langle \sigma_{j}^{+}(t)a_{in}^{R}(t-z_{N}/v_{g}) ] \nonumber\\
&+\sum_{jl}\frac{\sqrt{\Gamma_{i}\Gamma_{l}}}{2}e^{ik_{a}(z_{j}-z_{l})}\langle \sigma_{j}^{+}(t)\sigma_{l}^{-}(t)\rangle, \\
l(t)=& \langle a_{in}^{L+}(t+z_{1}/v_{g})a_{in}^{L}(t+z_{1}/v_{g})\rangle \nonumber\\
&-2 \sum_{j=1}^{N_a}\sqrt{\frac{\Gamma_{j}}{2}}\text{Im}[e^{-i k_{a}z_{j}}\langle \sigma_{j}^{+}(t)a_{in}^{L}(t+z_{1}/v_{g}) ] \nonumber\\
&+\sum_{jl}\frac{\sqrt{\Gamma_{i}\Gamma_{l}}}{2}e^{ik_{a}(z_{l}-z_{j})}\langle \sigma_{j}^{+}(t)\sigma_{l}^{-}(t)\rangle. 
\end{align}
On the right hand side of Eqs. (22) and (23),  the first terms are the incident field intensities,  the second terms are the absorption and stimulated emission of the system, and the last terms are the spontaneous emission of the system.  
From $r(t)$ and $l(t)$, we can obtain the pulse shape propagating to the right and to the left after the scattering process.
The field intensity reflected to the left and the right in the whole scattering process are then given by $I_R=\int_{0}^{\infty}r(t) dt$ and $I_L=\int_{0}^{\infty}l(t)dt$. Supposing that the photon pulse is initially propagating to the right, the reflectivity of the pulse  is then given by
\begin{equation}
R=\frac{I_{L}}{I_{R}+I_{L}},
\end{equation}
and the transmissivity $T=1-R$.  

The scattering power spectrum can  be usually obtained from the two-time correlation function of the output field operator
\begin{equation}
S(\omega)=\int_{0}^{\infty}\int_{0}^{\infty}\langle a_{out}^{+}(t_{1})a_{out}(t_{2})\rangle e^{i\omega(t_{1}-t_{2})}dt_{1}dt_{2},
\end{equation}
where the average is over the initial state of the whole system. According to the generalized  input-output relation shown in Eqs. (20) and (21),  $a_{out}(t)$ can be expressed as the summation of the input field operator  $a_{in}(t)$ and the emitter operators $\sigma_{j}^{-}(t)$. The results when $a_{in}(t)$ operator acts on the initial state can be readily worked out. Usually, the two-time average of the emitter operators $\langle \sigma_{j}^{+}(t)\sigma_{l}^{-}(t+\tau) \rangle$ can be calculated from the master equation according to the quantum regression theorem \cite{Scully1997}. 
However, to apply the quantum regression theorem to  calculate the two-time correlation function, it usually requires that the reservoir field does not change significantly. This condition may not be very well satisfied in the waveguide-QED system because the waveguide photon can be significantly absorbed by the emitters especially near the resonance frequency. Therefore, direct use of the quantum regression theorem to numerically calculate the spectrum here may cause some errors and need to be treated carefully. However, at the plane wave limit,  an alternative strategy can be used to calculate the scattering property of the system. If the incident photon pulse has a very narrow bandwidth, we can calculate its reflectivity and transmissivity from the above discussions. Repeating these procedures for each incident frequency, we can then obtain the scattering property of the waveguide-QED system  at the plane wave limit. However, we should note that the general power spectrum of this system driven by a pulse with finite bandwidth can not be completely captured by this frequency-sweep strategy and it should be calculated from Eq. (25).

\section{Application to different photon wavepackets}

In this section, we take the coherent states and the Fock states as example to show how to apply the theory we developed in the previous section to study the emitter dynamics and the field scattering property.

\subsection{Coherent state wavepacket}

We first consider the case when the incident field is a coherent photon pulse. Suppose that the incident field is a continuous-mode coherent state describing by wavefunction $|\Psi_{cs}\rangle =\Pi _k |\alpha_{k}\rangle$ where
\begin{equation}
 |\alpha_{k}\rangle =e^{-|\alpha_{k}|^2/2}\sum_{n_{k}=0}^{\infty}\frac{(\alpha_{k})^{n_k}}{\sqrt{n_{k}!}}|n_{k}\rangle.  
 \end{equation}
 The average photon for the $k$th mode $\bar{n}_{k}=|\alpha_{k}|^{2}$.
Since $a_{k}|\Psi_{cs}\rangle=\alpha_k |\Psi_{cs}\rangle$,  the operator $\rho'_{j}(t)=Tr_{R}[U(t)a_{j}(t)\rho(0) U^{\dagger}(t)]=\alpha_{j}(t)\rho_S(t)$  where 
\begin{equation} 
\alpha_{j}(t)=\sqrt{\frac{v_{g}}{2\pi}}\int_{-\infty}^{\infty} e^{ikz_{j}}e^{-i\delta\omega_{k}t}\alpha_{k}dk
\end{equation}
describes the real-time evolution of the incident coherent photon pulse. Therefore, the operator  $\rho'_{j}(t)$ is reduced to a number multiplying the system density operator $\rho_S(t)$.
The master equation shown in Eq. (13) then becomes
\begin{align}
\dot{\rho}_S(t)=&-\frac{i}{2}\sum_{j=1}^{N_{a}}\varepsilon_{j}(t)[\sigma_{j}^{z},\rho_{S}(t)] \nonumber\\
&-i\sum_{j=1}^{N_a}\sqrt{\frac{\Gamma_{j}}{2}}[\alpha_{j}(t)\sigma_{j}^{+}+\alpha_{j}^{*}(t)\sigma_{j}^{-},\rho_S(t)] \nonumber\\
&+i\sum_{jl}\text{Im}(\Lambda_{jl})[\rho_{S}(t),\sigma_j^{+}\sigma_l^{-}]+\mathcal{L}[\rho_{S}(t)],
\end{align}
where
$\mathcal{L}[\rho_{S}(t)]=-\sum_{jl}\text{Re}(\Lambda_{jl})[\sigma_{j}^{+}\sigma_{l}^{-}\rho_{S}(t)+\rho_{S}(t)\sigma_{j}^{+}\sigma_{l}^{-}-2\sigma_{l}^{-}\rho_{S}(t)\sigma_{j}^{+}]$
describes the collective dissipation process. Equation (28) is a general master equation of the waveguide-QED system when the incident photon pulse is in a coherent state and it reduces to the results shown in Ref. \cite{Shi2015} when there is not frequency modulation.

\begin{figure*}
	\includegraphics[width=0.6\columnwidth]{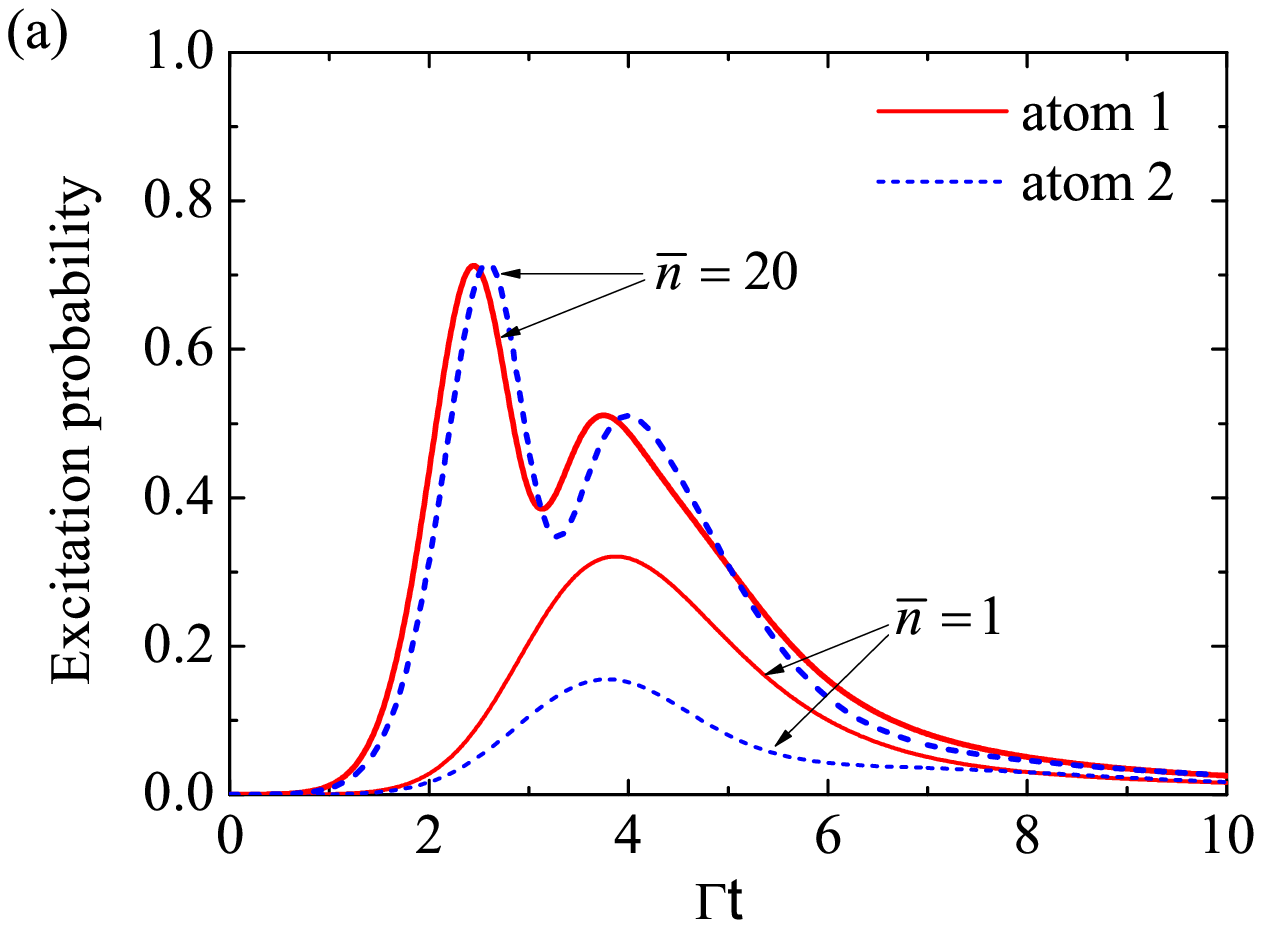}
	\includegraphics[width=0.6\columnwidth]{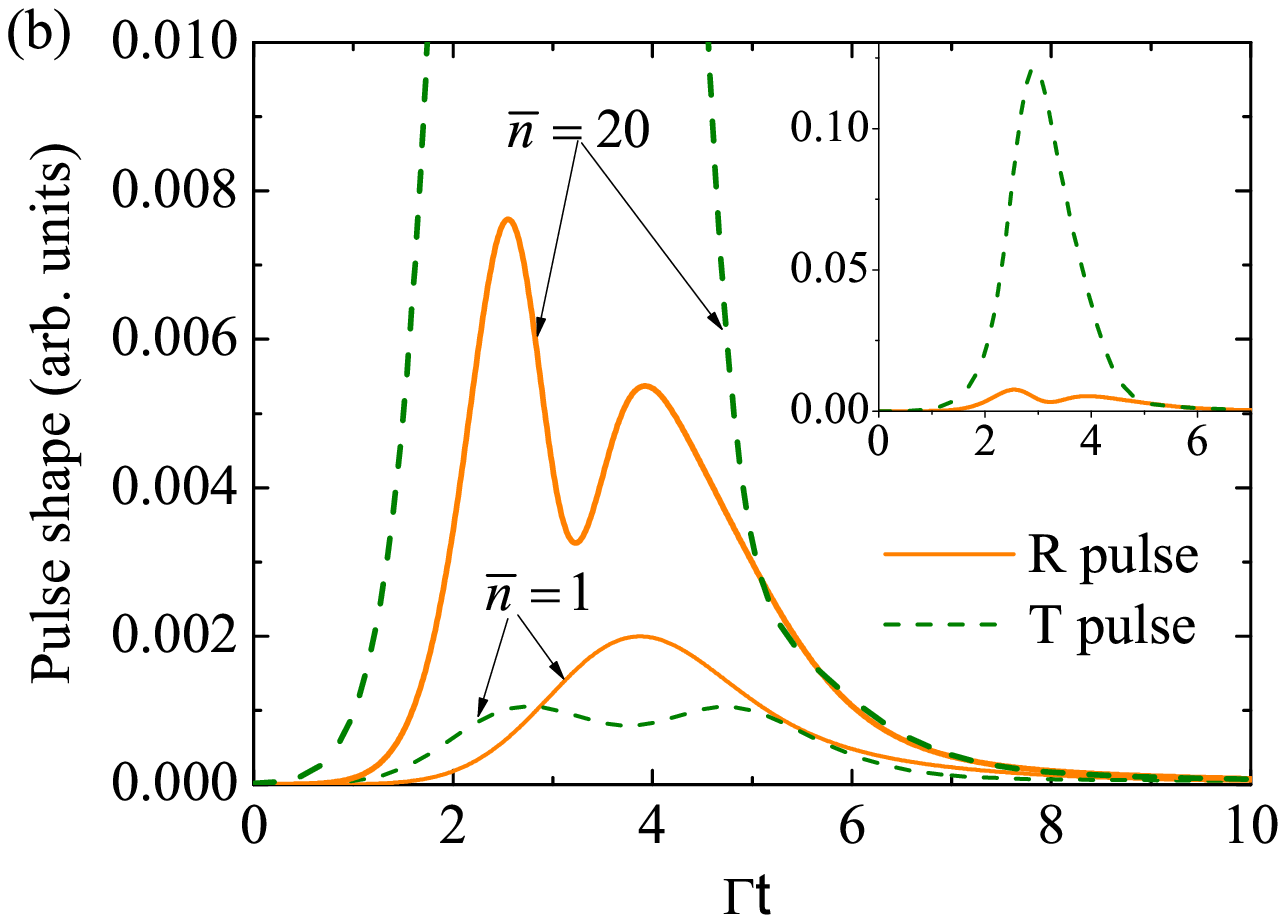}
	\includegraphics[width=0.6\columnwidth]{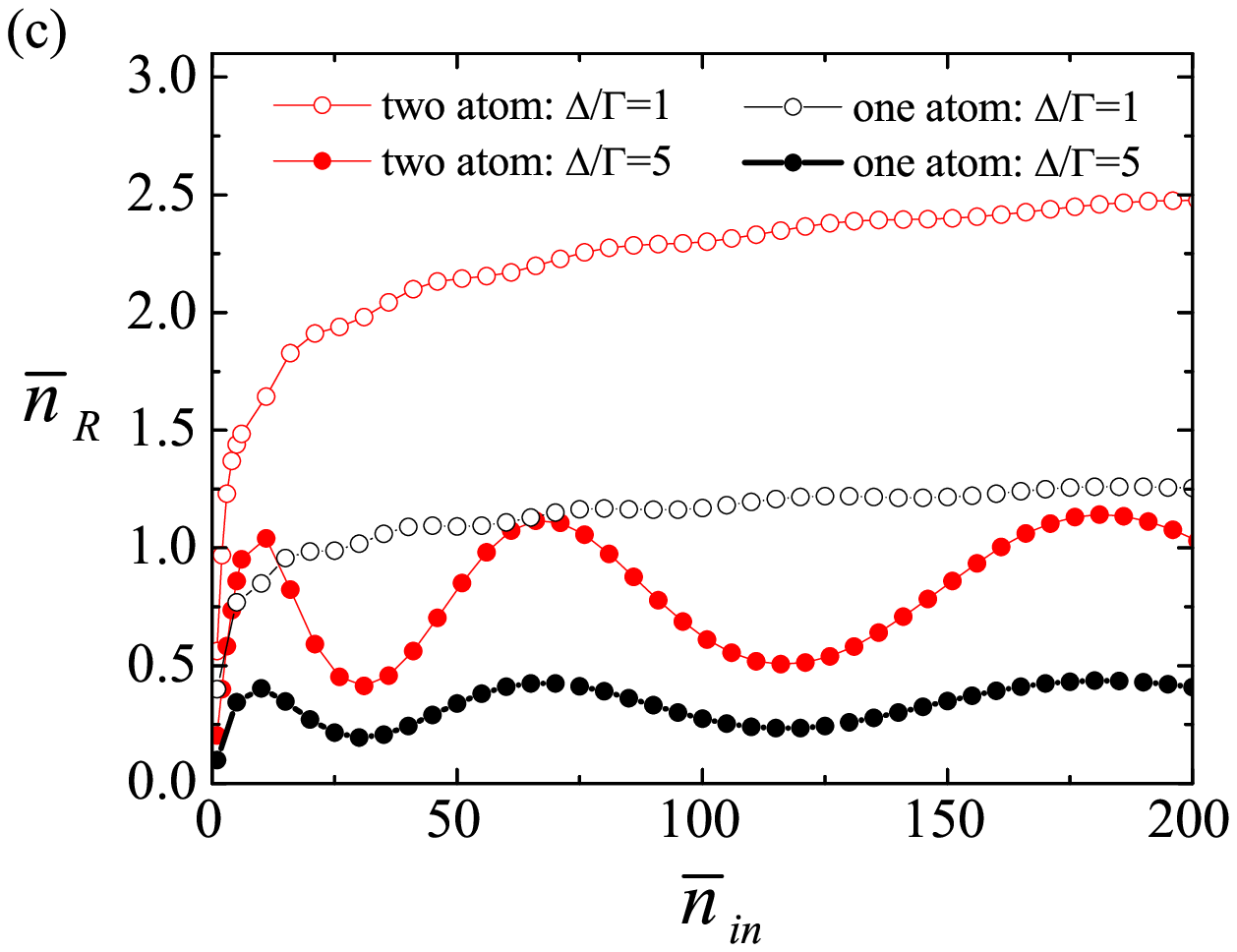}
	\caption{Coherent state pulse interacting with two emitters. (a) emitter excitation as a function of time for two different average photon number ($\bar{n}=1$ and $\bar{n}=20$). $\Delta v_{g}/\Gamma=1$. (b) Reflected and transmitted pulse shapes after the scattering for the same parameters as (a). (c) The average reflected photon number as a function of average incident photon number for two different pulse width ($\Delta/\Gamma=1$ and $\Delta/\Gamma=5$). For comparison, the results for one emitter are also shown as the black lines with solid ($\Delta/\Gamma=5$) and open circles ($\Delta/\Gamma=1$) .  The emitter distance $d=0.125\lambda_{a}$ for all three figures. }
	\label{2}
\end{figure*}

The master equation shown in Eq. (28) is itself a closed equation from which we can calculate the real-time dynamics of the emitters for arbitrary coherent pulse input. Our theory can be applied to calculate the dynamics of the system with arbitrary photon pulse shapes. Here, without loss of generality we assume that the photon pulse has a Gaussian shape throughout this paper. Supposing that the incident coherent field has a Gaussian pulse shape with average photon number $\bar{n} $, its spectrum can be written as
\begin{equation}
\alpha_{k}=\frac{\sqrt{\bar{n}}}{\pi^{1/4}\sqrt{\Delta}}e^{-(k-k_0)^2/2\Delta^2}e^{-ikz_{0}},
\end{equation}
where $z_0$ is the initial central peak position of the pulse and $k_{0}$ is the wavevector corresponding to the central frequency of the photon pulse. When $k_{0}>0$ ($k_{0}<0$) the pulse is propagating to the right (left). 
The average photon number is given by $\bar{n}=\sum_{k}\bar{n}_{k}=\int_{-\infty}^{\infty}|\alpha_{k}|^2 dk$.
For the right propagating incident pulse (i.e., $k_{0}>0$), we have
\begin{equation}
\alpha_{j}^{R}(t)=\frac{\sqrt{\bar{n}\Delta v_{g}}}{\pi^{1/4}}e^{-\frac{\Delta ^{2}(z_{j0}-v_{g}t)^2}{2}}e^{ik_{a}z_{j0}}e^{i\Delta_{k}(z_{j0}-v_{g}t)}.
\end{equation}
 For the left propagating incident pulse (i.e., $k_{0}<0$), we then have
\begin{equation}
\alpha_{j}^{L}(t)=\frac{\sqrt{\bar{n}\Delta v_{g}}}{\pi^{1/4}}e^{-\frac{\Delta ^{2}(z_{j0}+v_{g}t)^2}{2}}e^{-ik_{a}z_{j0}}e^{-i\Delta_{k}(z_{j0}+v_{g}t)},
\end{equation}
where $z_{j0}=z_{j}-z_{0}$ and $\Delta_{k}=|k_0|-k_{a}$ is the detuning between the center frequency of the pulse and the emitter transition frequency.

The numerical results for the coherent state input are shown in Fig. 2 where the coherent state is scattered by two emitters. We assume that the  distance between these two emitters is  $0.125\lambda_{a}$ where $\lambda_{a}=2\pi/k_{a}$. The excitations of the two emitters as a function of time for two different incident average photon number ($\bar{n}=1$ and $\bar{n}=30$) are shown in Fig. 2(a). When the average incident photon number is small, e.g. $\bar{n}=1$, both emitters are first excited and then deexcited as the coherent pulse passing through. However, when the average incident photon number is large, e.g. $\bar{n}=20$, the excitations of both emitters can have multiple peaks which is the signature of Rabi oscillations.

The corresponding reflected and transmitted photon pulse shapes after the scattering are shown in Fig. 2(b). When the average photon number is small, the reflected pulse (thinner orange solid line)  has a single peak and the transmitted pulse (thinner olive dashed line) has two peaks due to the interference between the incident photon wavefunction and the reemitted photon wavefunction.   When the average photon number is large, most photons are transmitted (thicker olive dashed line) and only a very small part of the photons are reflected (thicker orange solid line). This is because the pulse with large photon number can saturate the emitter excitation quickly and only a very small part of photons can be absorbed. Here, the reflection photon pulse can have two peaks instead of one peak  due to the Rabi oscillations which does not occur when the photon number is small. 

For a coherent pulse with finite time duration, the average photon number reflected by the emitters may be saturated. Here, we also study the average reflected photon number $\bar{n}_{R}$ as a function of average incident photon number $\bar{n}_{in}$ for two fixed pulse spectrum widths ($\Delta=\Gamma$ and $\Delta=\Gamma/5$ ) and the results are shown in Fig. 2(c) when the distance between the two emitters is $0.125\lambda_{a}$.  When the pulse width is about $\Gamma$, the average reflected photon number increases quickly first  as $\bar{n}_{in}$ increases but then it increases extremely slowly when $\bar{n}_{in}$ is large due to the saturation effect (red line with open circles in Fig. 2(c)). It is also noted that when the incident photon number is large, the average reflected photon number can be larger than two despite that there are only two emitters. This is because the incident pulse is not short enough to saturate the emitters immediately.  When the incident pulse duration is shorter, i.e. the incident pulse has a broader spectrum (e.g.,   $\Delta=5\Gamma$), $\bar{n}_{R}$ first increases and then oscillates as $\bar{n}_{in}$ increases (red line with solid circles in Fig. 2(c)) due to the stimulated emission effects. The average reflected photon number is obviously less than 2 because the shorter pulse can saturate the emitters quickly. For comparison, we also plot the results when there is only a single emitter in the system (black lines with open and closed circles). We can see that their behaviors are similar but the average reflected photon number for two emitters is larger than that of the single emitter. When the pulse duration is much smaller than the decay time of the emitter, the average photon number being reflected by a single emitter is always less than one which can be exploited to produce single photon sources \cite{Senellart2017}.

\subsection{Single photon wavepacket}

\begin{figure*}
	\includegraphics[width=0.6\columnwidth]{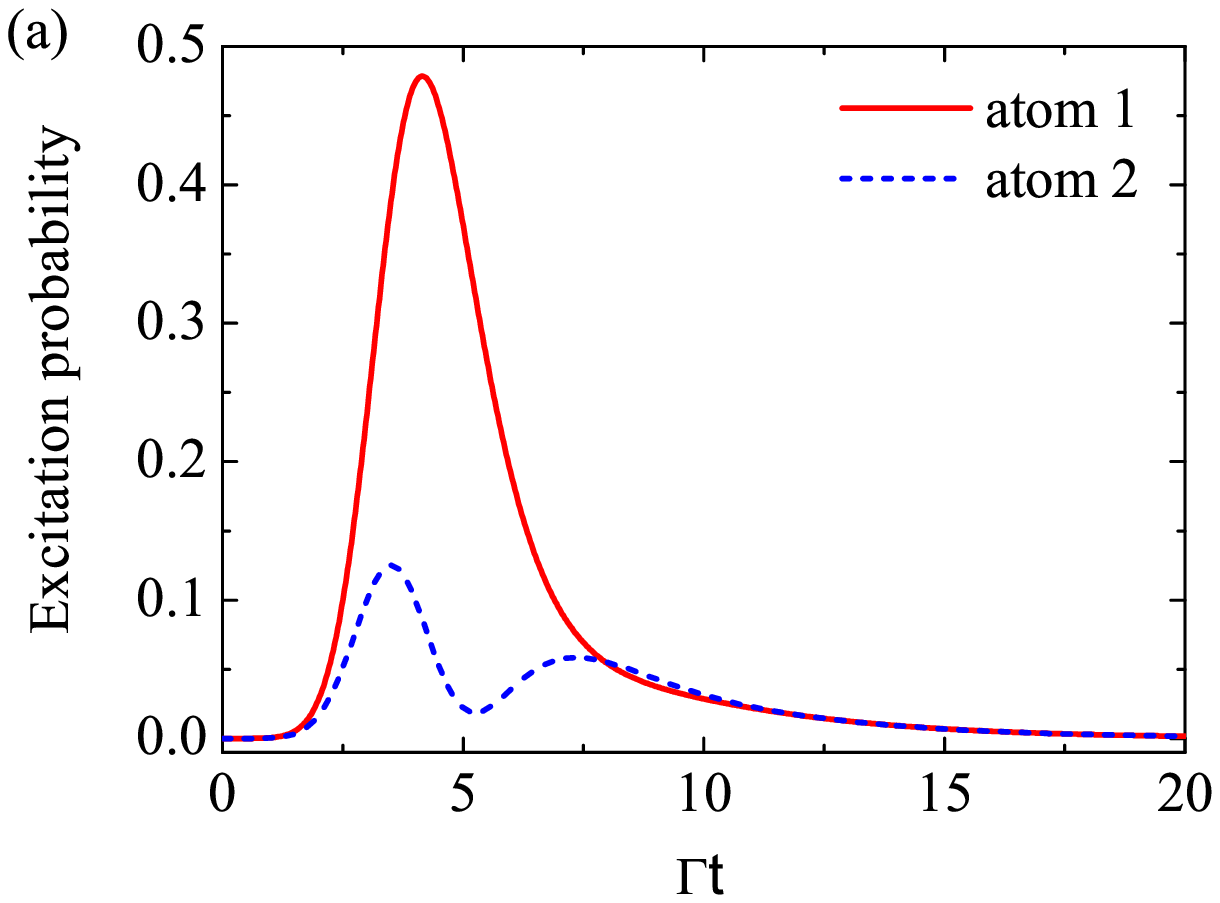}
	\includegraphics[width=0.6\columnwidth]{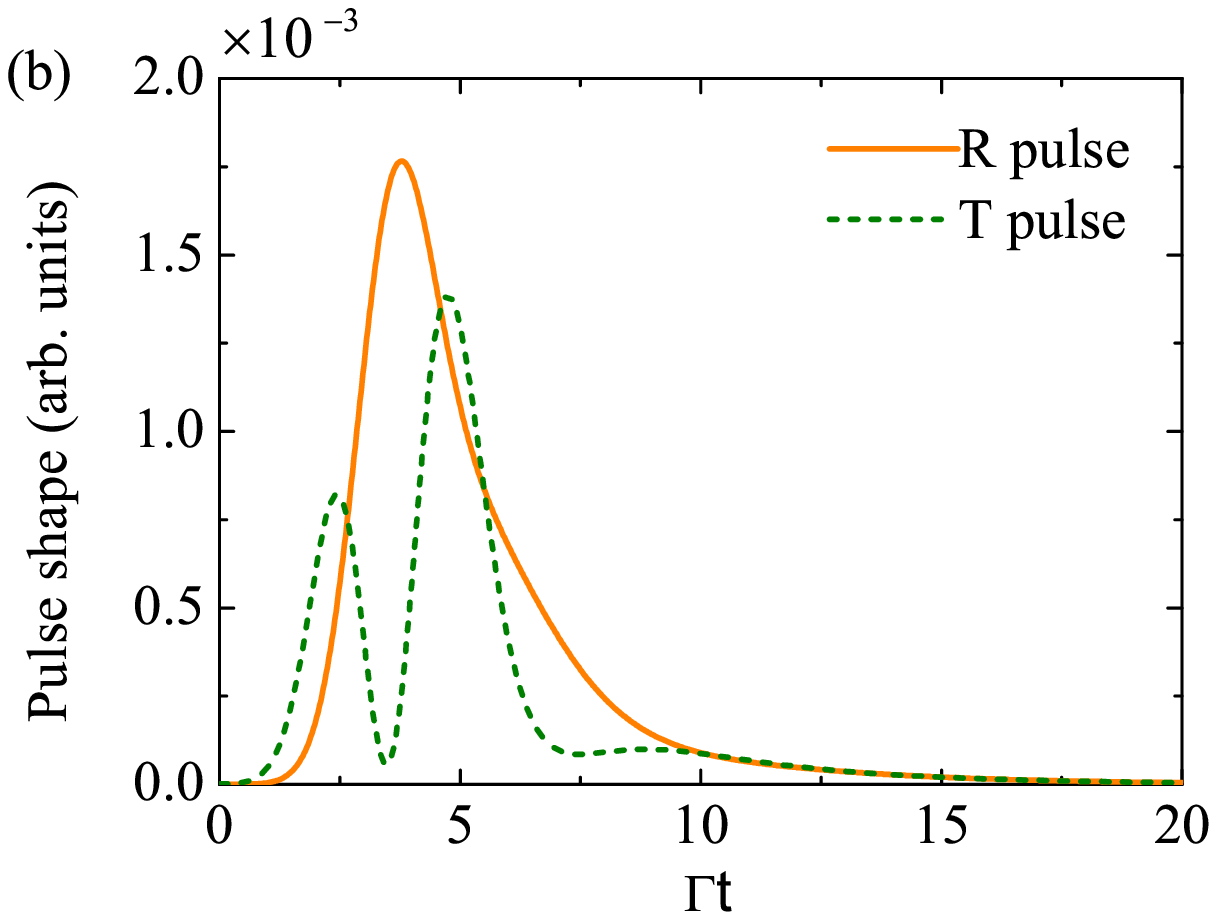}
	\includegraphics[width=0.6\columnwidth]{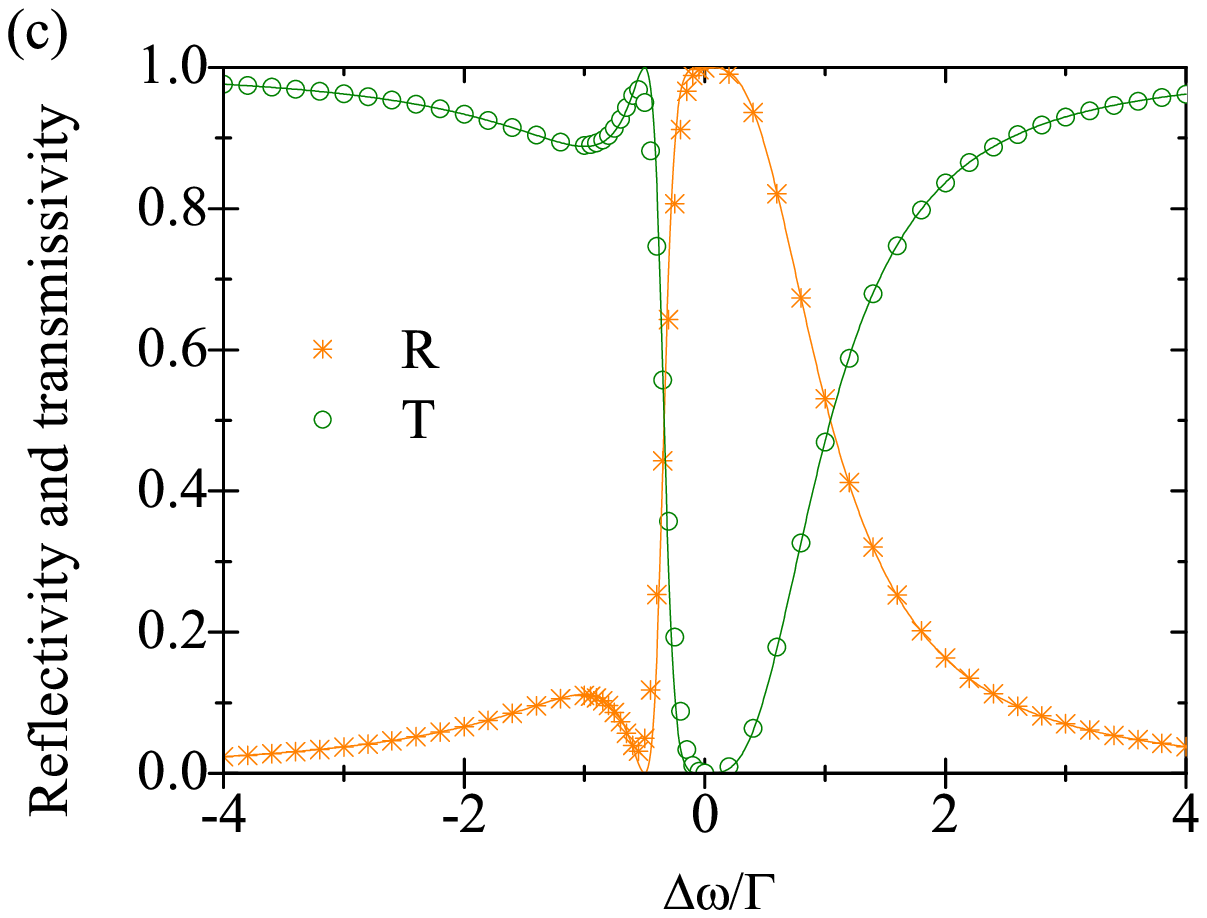}
	\caption{A single photon wavepacket interacting with two emitters. (a) emitter excitation as a function of time. $\Delta/\Gamma=1$. (b) Reflected and transmitted pulse shapes after the scattering for the same parameters as (a).  (c) The reflectivity and transmissivity as a function of photon frequency where the symbols are the numerical results and the solid lines are the analytical results. In all three figures, $d=0.125\lambda$.}
	\label{3}
\end{figure*}

Compared with the coherent state input, the calculation of Fock state input is more involved mostly because of its quantum nature. The theory developed in Sec. II can also be applied for the arbitrary Fock state inputs. In this subsection, we consider the simplest case where the pulse only contains a single photon. Actually, for the single photon pulse case, we have developed a dynamical transport theory for calculating the real-time evolution of the system based on the wavefunction approach \cite{Liao2015, Liao2016}. Here, we instead use the master equation developed in Sec. II to calculate the system dynamics.

The single photon wavepacket can be described by the wavefunction
 \begin{equation} 
 |\Psi_{F}\rangle = \int_{-\infty}^{\infty} \alpha (k) a^{\dagger}_{k}|0\rangle dk,
 \end{equation}
 where $\int_{-\infty}^{\infty} |\alpha (k)|^{2} dk =1$. Since $a_k|\Psi_{F}\rangle=\alpha_{k}|0\rangle $, the $\rho'_{j}(t)$ term in Eq. (13) is then given by 
 \begin{equation}
 \rho'_{j}(t)=Tr_{R}[U(t)a_{j}(t)\rho(0)U^{+}(t)]=\alpha_{j}(t)\rho_{01}^{S}(t),
 \end{equation} 
where $\alpha_{j}(t)$ is given by Eq. (27) and we define a new operator $\rho_{01}^{S}(t)=Tr_{R}[U(t)\rho_{S}(0)\otimes |0\rangle \langle\Psi_{F}| U^{+}(t)]$. 
If we define $\rho_{11}^{S}(t)=Tr_{R}[U(t)\rho^{S}(0)\otimes |\Psi_F\rangle \langle \Psi_{F}| U^{\dagger}(t)]$, we can obtain from Eq. (13) that
\begin{align}
\dot{\rho}^{S}_{11}(t)=&-\frac{i}{2}\sum_{j=1}^{N_{a}}\varepsilon_{j}(t)[\sigma_{j}^{z},\rho^{S}_{11}(t)] \nonumber\\ &-i\sum_{j=1}^{N_a}\sqrt{\frac{\Gamma_{j}}{2}}[\alpha_{j}(t)\sigma_{j}^{+}(t),\rho_{01}^{S}(t) ] \nonumber\\
&-i\sum_{j=1}^{N_a}\sqrt{\frac{\Gamma_{j}}{2}}[\alpha_{j}^{*}(t)\sigma_{j}^{-}(t),\rho_{01}^{S\dagger}(t)] \nonumber \\
&-i\sum_{jl}\text{Im}(\Lambda_{jl}) [\sigma_{j}^{+}\sigma_{l}^{-},\rho^{S}_{11}(t)] -\mathcal{L}[\rho^{S}_{11}(t)],
\end{align}
where $\mathcal{L}[\rho^{S}_{11}(t)]=\sum_{jl}\text{Re}(\Lambda_{jl}) [\sigma_{j}^{+}\sigma_{l}^{-}\rho^{S}_{11}(t)+\rho^{S}_{11}(t)\sigma_{j}^{+}\sigma_{l}^{-}-2\sigma_{l}^{-}\rho^{S}_{11}(t)\sigma_{j}^{+}]$ is the collective dissipation term.  $\rho_{01}^{S}(t)$ is not a valid density matrix because it is traceless but it satisfies  $\rho_{01}^{S\dagger}=\rho_{10}^{S}$. Since a new  operator $\rho_{01}^{S}(t)$ appears, Eq. (34) is itself not a closed equation and we need to derive an extra equation for  $\rho_{01}^{S}(t)$.

The dynamical equation for $\rho_{01}(t)$ can be derived using similar procedures as deriving $\rho_{S}(t)$ shown in Sec. II and it is given by (see Appendix C)
\begin{align}
\dot{\rho}_{01}^{S}(t)=&-\frac{i}{2}\sum_{j=1}^{N_{a}}\varepsilon_{j}(t)[\sigma_{j}^{z},\rho_{01}^{S}(t)] \nonumber\\ &-i\sum_{j=1}^{N_a}\sqrt{\frac{\Gamma_{j}}{2}}\alpha_{j}^{*}(t) [\sigma_{j}^{-},\rho_{00}^{S}(t)] \nonumber\\
&-i\sum_{jl}\text{Im}(\Lambda_{jl}) [\sigma_{j}^{+}\sigma_{l}^{-},\rho_{01}^{S}(t)] -\mathcal{L}[\rho_{01}^{S}(t)],
\end{align}
where $\rho_{00}^{S}(t)=Tr_{R}[U(t)\rho_S \otimes |0\rangle \langle 0|U^{\dagger}(t)]$ is another density matrix describing the evolution of the system when  the field is initially in the vacuum. Using similar procedure, it is not difficult to obtain that 
\begin{align}
\dot{\rho}^{S}_{00}(t)=&-\frac{i}{2}\sum_{j=1}^{N_{a}}\varepsilon_{j}(t)[\sigma_{j}^{z},\rho^{S}_{00}(t)] \nonumber\\
&-i\sum_{jl}\text{Im}(\Lambda_{jl}) [\sigma_{j}^{+}\sigma_{l}^{-},\rho^{S}_{00}(t)]-\mathcal{L}[\rho^{S}_{00}(t)],
\end{align}
where we see that no new density operator appears. 

Hence, the master equation for the single photon state input consists of three cascaded equations as given by Eqs. (34-36) while only a single equation is needed in the coherent state input. The dynamics of the emitters for arbitrary single photon pulse input can then be calculated from these three equations.  The time evolution of the average value of an arbitrary emitter operator  $O(t)$ can be calculated as $\langle O(t) \rangle = Tr_{S}[O\rho_{11}^{S}(t)]$.

One numerical example is shown in Fig. 3 where we consider a single photon wavepacket interacting with two emitters. Here, we assume that the single photon wavepacket has a Gaussian spectrum as shown in Eq. (29) and the distance between emitters  is $\lambda_{a}/8$. The emitter excitation as a function of time is shown in Fig. 3(a). Due to the collective interaction, the first emitter can have much higher excitation probability than that of the second one and the excitation of the second emitter has a Rabi-like oscillations which does not occur when the incident photon pulse is in a coherent state with $\bar{n}_{in}=1$. This is due to the interference between the two excitation channels, i.e., the excitation of the incident photon and the excitation by the first excited emitter. In the coherent state input, this interference is however concealed.

The reflected and transmitted photon pulse shapes are shown in Fig. 3(b) from which we can see that the transmitted pulse has multiple peaks due to the quantum interference between the incident photon and the reemitted photons by the two emitters. The visibility of the oscillation is much larger than that in the coherent state input. The emitter dynamics and the scattering pulse shapes shown in Figs. 3(a) and 3(b) are the same as those calculated by the wavefunction approach \cite{Liao2015}.  The reflected and transmitted spectra when the incident single photon is a plane wave are shown in  Fig. 3(c) where the symbols are the numerical results and the solid lines are the analytical results calculated by the stationary scattering theory such as the Bethe-ansatz approach \cite{Shen2005}. It clearly shows that for single photon input the results calculated by our theory here are consistent with those shown in the previous literatures.  The spectrum shown in Fig. 3(c)  clearly shows an asymmetric Fano-like structure. This is caused by the interference between the two collective emission channels, i.e., the emission from the two collective excited states $|\pm\rangle=\frac{1}{\sqrt{2}}(|eg\rangle \pm |ge\rangle )$ which have different energy shifts and decay rates.  The results shown in this subsection confirm the validity of our theory developed here.

\subsection{N-photon wavepacket}

In addition to the single photon Fock state, we can also derive generalized master equations for the multi-photon Fock state inputs. Compared with the single-photon input, the calculation of multi-photon Fock state input is more complicated. We first consider a relative simple subset which is the direct generalization of the single photon wavepacket, i.e.,  
\begin{equation}
|N_\alpha\rangle=\frac{1}{\sqrt{N!}}\Big[\int_{-\infty}^{\infty} dk \alpha(k)a_{k}^{\dagger}\Big]^{N}|0\rangle,
\end{equation} 
where we have the normalization condition $\int_{-\infty}^{\infty} |\alpha_{k}|^{2}dk=1$ \cite{Baragiola2012}. A general N-photon wavepacket can be always decomposed into the superposition of the wavefunction shown in Eq. (37) and we have
\begin{equation}
a_{k}|N_\alpha\rangle=\sqrt{N}\alpha(k)|N-1_{\alpha}\rangle.
\end{equation}
In general, we have the relation $a_{k}|m_{\alpha}\rangle=\sqrt{m}\alpha(k)|m-1_\alpha\rangle$ and therefore
\begin{equation}
a_{k}\rho_{s}(0)\otimes |m_{\alpha}\rangle \langle n_{\alpha}|=\sqrt{m}\alpha(k)\rho_{s}(0)\otimes |m-1_{\alpha}\rangle \langle n_{\alpha}|.
\end{equation}

Using the similar procedures to derive Eq. (13), we can derive a ladder set of dynamical equations for the N-photon wavepacket input which is given by
\begin{widetext}
\begin{align}
\dot{\rho}_{mn}^{S}(t)=&-\frac{i}{2}\sum_{j=1}^{N_{a}}\varepsilon_{j}(t)[\sigma_{j}^{z},\rho_{mn}^{S}(t)]-i\sum_{j=1}^{N_a}\sqrt{\frac{\Gamma_{j}}{2}}\{\sqrt{m}\alpha_{j}(t)[\sigma_{j}^{+},\rho_{m-1,n}^{S}(t)] +\sqrt{n}\alpha_{j}^{*}(t)[\sigma_{j}^{-}(t),\rho_{mn-1}^{S}(t)]\}\nonumber \\  &-i\sum_{jl}\text{Im}(\Lambda_{jl})[\sigma_{j}^{+}\sigma_{l}^{-},\rho^{S}_{mn}(t)]-\mathcal{L}[\rho^{S}_{mn}(t)],
\end{align}
\end{widetext}
where $\rho_{mn}^{S}(t)=Tr_{R}[U(t)\rho_{S}(0)\otimes|m\rangle\langle n | U^{+}(t)]$ and $0\leq m,n \leq N$. Considering that $\rho_{mn}^{S}=\rho_{nm}^{S\dagger}$,  $(N+1)(N+2)/2$ master equations are required to make the equations closed where $N$ is the total incident photon number. For example, three master equations are needed for single-photon input which is shown in the previous subsection, while for two-photon input we need six cascaded master equations. 

Taking the two-photon input as an example,  the master equations are given by
\begin{widetext}
\begin{align}
\dot{\rho}_{22}^{S}(t)=&-\frac{i}{2}\sum_{j=1}^{N_{a}}\varepsilon_{j}(t)[\sigma_{j}^{z},\rho_{22}^{S}(t)] -i\sum_{j=1}^{N_a}\sqrt{\frac{\Gamma_{j}}{2}}\{\sqrt{2}\alpha_{j}(t)[\sigma_{j}^{+},\rho_{12}^{S}(t)] +\sqrt{2}\alpha_{j}^{*}(t)[\sigma_{j}^{-}(t),\rho_{21}^{S}(t)]\}\nonumber \\ &-i\sum_{jl}\text{Im}(\Lambda_{jl})[\sigma_{j}^{+}\sigma_{l}^{-},\rho^{S}_{22}(t)] -\mathcal{L}[\rho^{S}_{22}(t)],
\\
\dot{\rho}_{12}^{S}(t)=&-\frac{i}{2}\sum_{j=1}^{N_{a}}\varepsilon_{j}(t)[\sigma_{j}^{z},\rho_{12}^{S}(t)] -i\sum_{j=1}^{N_a}\sqrt{\frac{\Gamma_{j}}{2}}\{\alpha_{j}(t)[\sigma_{j}^{+},\rho_{02}^{S}(t)] +\sqrt{2}\alpha_{j}^{*}(t)[\sigma_{j}^{-}(t),\rho_{11}^{S}(t)]\}\nonumber \\ &-i\sum_{jl}\text{Im}(\Lambda_{jl})[\sigma_{j}^{+}\sigma_{l}^{-},\rho^{S}_{12}(t)] -\mathcal{L}[\rho^{S}_{12}(t)],
\\
\dot{\rho}_{11}^{S}(t)=&-\frac{i}{2}\sum_{j=1}^{N_{a}}\varepsilon_{j}(t)[\sigma_{j}^{z},\rho_{11}^{S}(t)] -i\sum_{j=1}^{N_a}\sqrt{\frac{\Gamma_{j}}{2}}\{\alpha_{j}(t)[\sigma_{j}^{+},\rho_{01}^{S}(t)] +\alpha_{j}^{*}(t)[\sigma_{j}^{-}(t),\rho_{10}^{S}(t)]\}\nonumber \\ &-i\sum_{jl}\text{Im}(\Lambda_{jl})[\sigma_{j}^{+}\sigma_{l}^{-},\rho^{S}_{11}(t)] -\mathcal{L}[\rho^{S}_{11}(t)],
\\
\dot{\rho}_{02}^{S}(t)=&-\frac{i}{2}\sum_{j=1}^{N_{a}}\varepsilon_{j}(t)[\sigma_{j}^{z},\rho_{02}^{S}(t)] -i\sum_{j=1}^{N_a}\sqrt{\frac{\Gamma_{j}}{2}}\{\sqrt{2}\alpha_{j}^{*}(t)[\sigma_{j}^{-}(t),\rho_{01}^{S}(t)]\} -i\sum_{jl}\text{Im}(\Lambda_{jl})[\sigma_{j}^{+}\sigma_{l}^{-},\rho^{S}_{02}(t)]  -\mathcal{L}[\rho^{S}_{02}(t)],
\\
\dot{\rho}_{01}^{S}(t)=&-\frac{i}{2}\sum_{j=1}^{N_{a}}\varepsilon_{j}(t)[\sigma_{j}^{z},\rho_{01}^{S}(t)] -i\sum_{j=1}^{N_a}\sqrt{\frac{\Gamma_{j}}{2}}\{\alpha_{j}^{*}(t)[\sigma_{j}^{-}(t),\rho_{00}^{S}(t)]\} -i\sum_{jl}\text{Im}(\Lambda_{jl})[\sigma_{j}^{+}\sigma_{l}^{-},\rho^{S}_{01}(t)] -\mathcal{L}[\rho^{S}_{01}(t)], \\
\dot{\rho}_{00}^{S}(t)=&-\frac{i}{2}\sum_{j=1}^{N_{a}}\varepsilon_{j}(t)[\sigma_{j}^{z},\rho_{00}^{S}(t)] -i\sum_{jl}\text{Im}(\Lambda_{jl})[\sigma_{j}^{+}\sigma_{l}^{-},\rho^{S}_{00}(t)] -\mathcal{L}[\rho^{S}_{00}(t)],
\end{align}
\end{widetext}
and we have $\rho_{nm}^{S}=\rho_{mn}^{S\dagger}$. Hence, for two-photon wavepacket, six cascaded master equations are required to calculate the dynamics of the system.

\begin{figure*}
	\includegraphics[width=0.6\columnwidth]{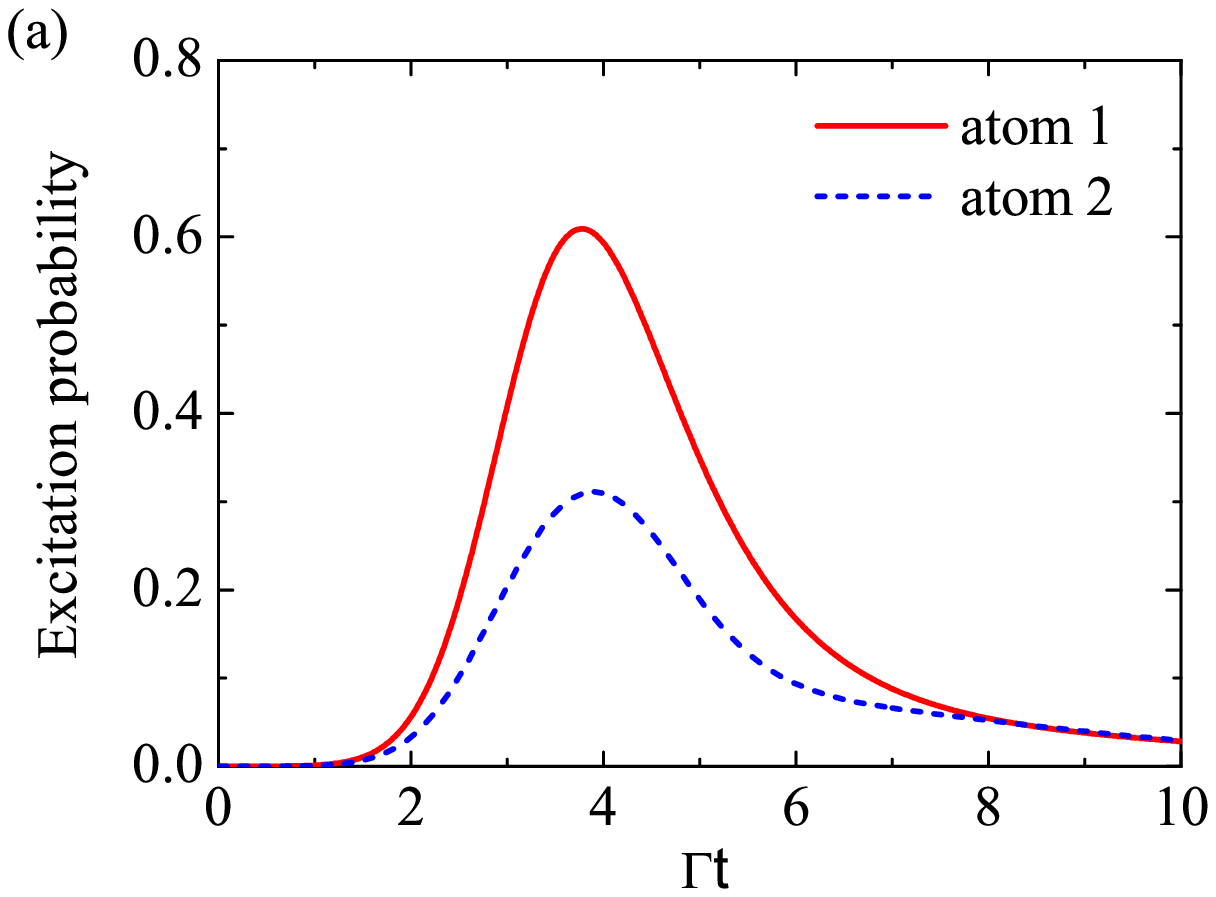}
	\includegraphics[width=0.6\columnwidth]{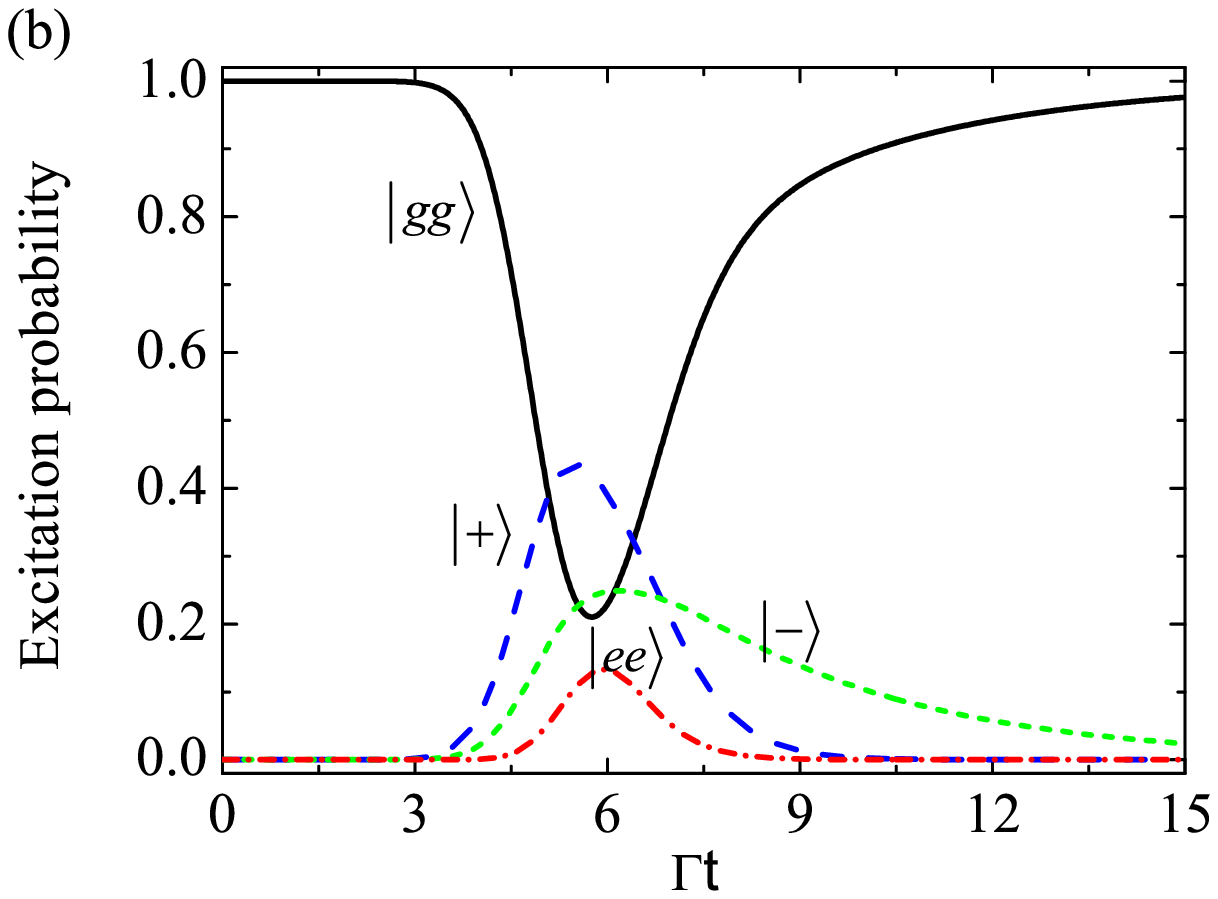}
	\includegraphics[width=0.6\columnwidth]{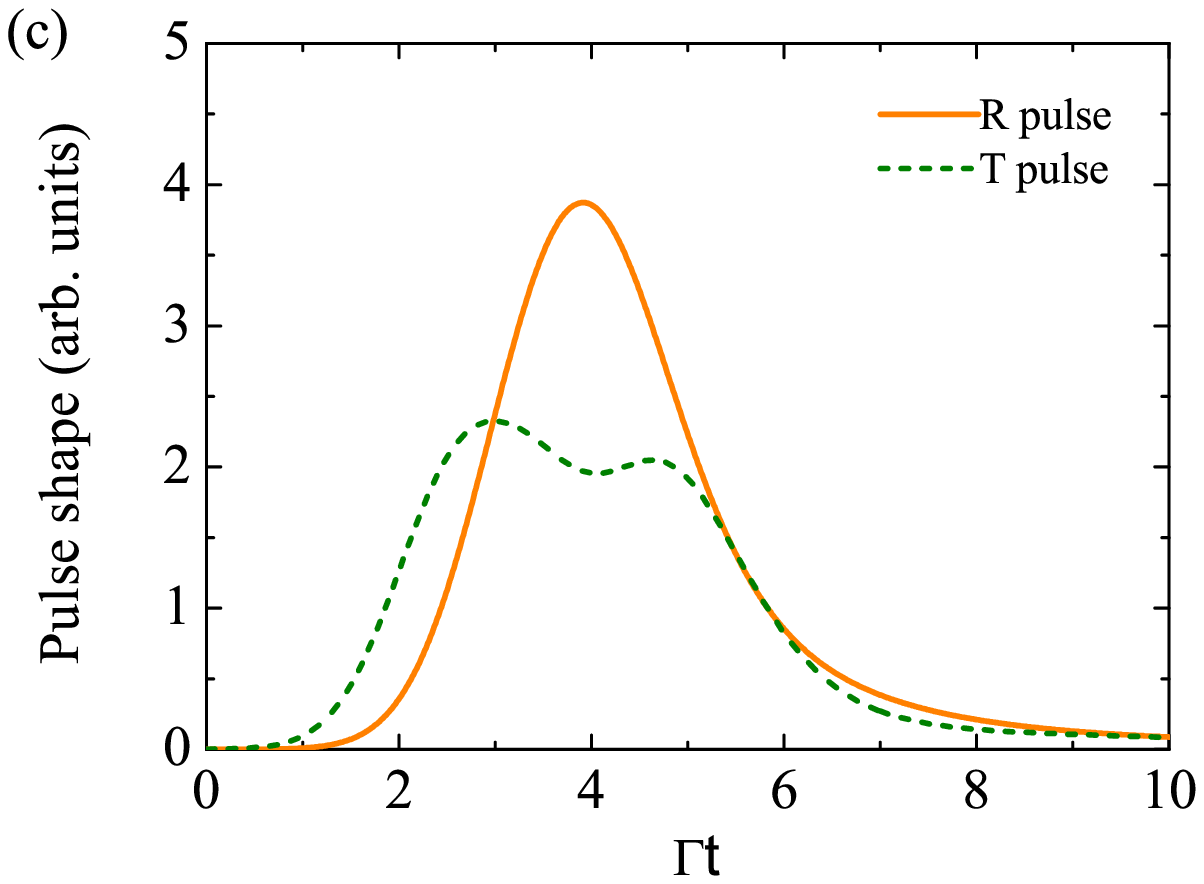}
	\caption{Two photon wavepacket interacting with two emitters. (a) Emitter excitation as a function of time. (b) The excitation of different eigenstates as a function of time.  (c) Reflected and transmitted pulse shapes after the scattering for the same parameters as (a).   In both figures, $\lambda_{a}/8$ and  $\Delta v_{g}/\Gamma=1$.}
	\label{4}
\end{figure*}

\begin{figure}
	\includegraphics[width=0.8\columnwidth]{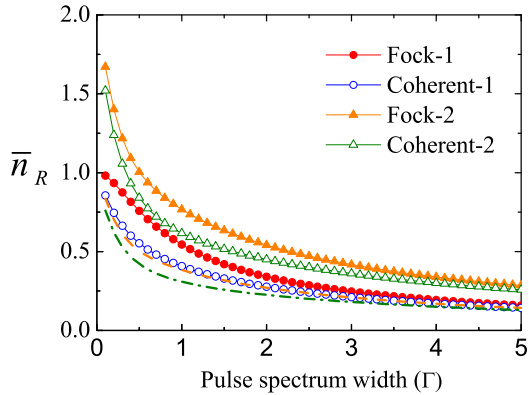}
	\caption{The average reflected photon number by a single emitter as a function of pulse width for four different incident photon states. Fock-1: a single photon state; Fock-2: two photon state; Coherent-1: coherent state with 1 average photon number; Coherent-2: coherent state with 2 average photon number. The orange dashed line is the corresponding reflectivity for the two photon state. The olive dashed-dotted line is the corresponding reflectivity of the Coherent-2 state. }
	\label{5}
\end{figure}

\begin{figure}
	\includegraphics[width=0.49\columnwidth]{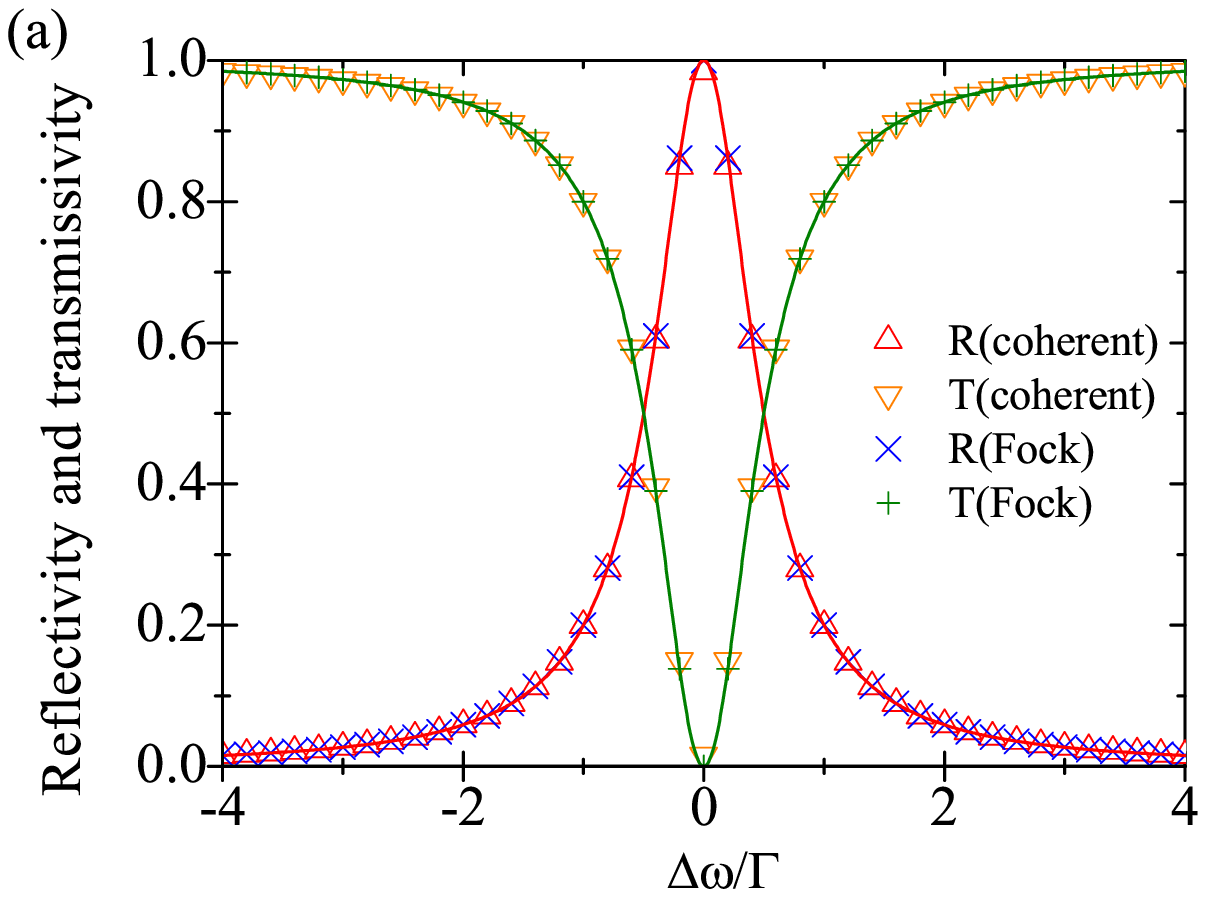}
	\includegraphics[width=0.49\columnwidth]{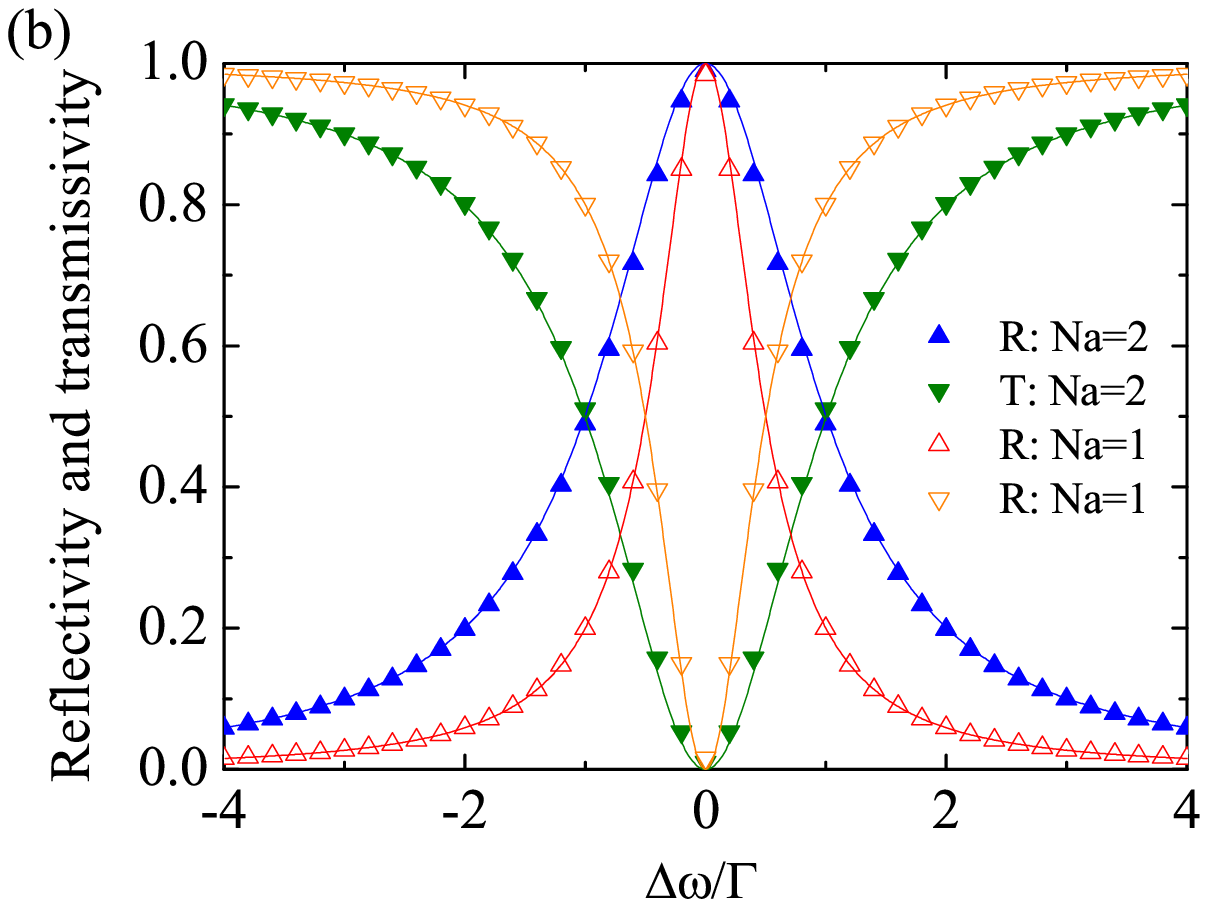}
	\caption{The reflectivity and transmissivity at the plane wave limit. (a)  A single emitter case for coherent state input and Fock state input. (b) Comparison of a single-emitter case and two-emitter case with separation $0.5\lambda_{a}$.  }
	\label{5}
\end{figure}

A numerical example is shown in Fig. 4 where we consider two-photon interacting with two emitters. Similar to the single-photon case,  we also assume that the two-photon pulse has a Gaussian spectrum and the distance between the emitters is $\lambda_{a}/8$. Compared with the single photon case, the emitter excitation in the two-photon case is larger and both excitations increase first and then decrease which is somewhat similar to the coherent state input (Fig. 4(a)). Different from the single photon case, the emitter 2 does not have Rabi-oscillation like structure. This is mainly because the double excited state $|ee\rangle$ can  also be populated in the two photon cases (red dashed-dotted line in Fig. 4(b)) and it can cover the interference effect which occurs in the single photon case. From Fig. 4(b), we can also see that the subradiant state $|-\rangle$ can be populated and it can last for extended period of time (green short dashed line). The symmetric state $|+\rangle$ is excited and deexcited much faster than the subradiant state $|-\rangle$ and it is a superradiant state (blue dashed line). The corresponding reflected and transmitted pulse shapes are shown in Fig. 4 (c) from which we can see that they are similar to those in the single photon case but the transmitted pulse has only a small oscillation in the two photon case. 

\subsection{The effects of pulse spectrum width}

In the stationary scattering theory, the incident field is usually assumed to be a plane wave. In practical experiments, the incident light is always a pulse with finite time duration and finite spectrum bandwidth. Here, our theory allows us to study the effects of the pulse widths. 

Taking the single emitter as an example, we investigate the average reflected photon number as a function of pulse spectrum widths for different input photon states. The results are shown in Fig. 5. For all four incident pulses, the average reflected photon number $\bar{n}_{R}$ decreases when the pulse spectrum width increases (i.e., the pulse time duration becomes shorter). This is because of the saturation effects. When the pulse is short, it can quickly saturate the emitter absorption and therefor the average reflected photon number decreases. When the pulse has a white spectrum (i.e, the pulse duration is extremely short), almost no photon will be reflected for both the coherent  state inputs and the Fock state inputs because most photons have frequencies far detuned from the resonance frequency of the emitter. In contrast,  when the pulse spectrum is extremely narrow (i.e., the pulse is at the plane wave limit) and its frequency is in resonance with the emitter transition frequency, almost all of the incident photons will be reflected for both the Fock state inputs and the coherent state inputs. When the pulse spectrum width is finite, the Fock state input can have larger reflectivity than that of the coherent state input with the same average incident photon number. For the same pulse width, the  pulse with $\bar{n}_{in}=1$ has larger reflectivity than that of  the pulse with $\bar{n}_{in}=2$ due to saturation effects. 

\begin{figure*}
	\includegraphics[width=0.6\columnwidth]{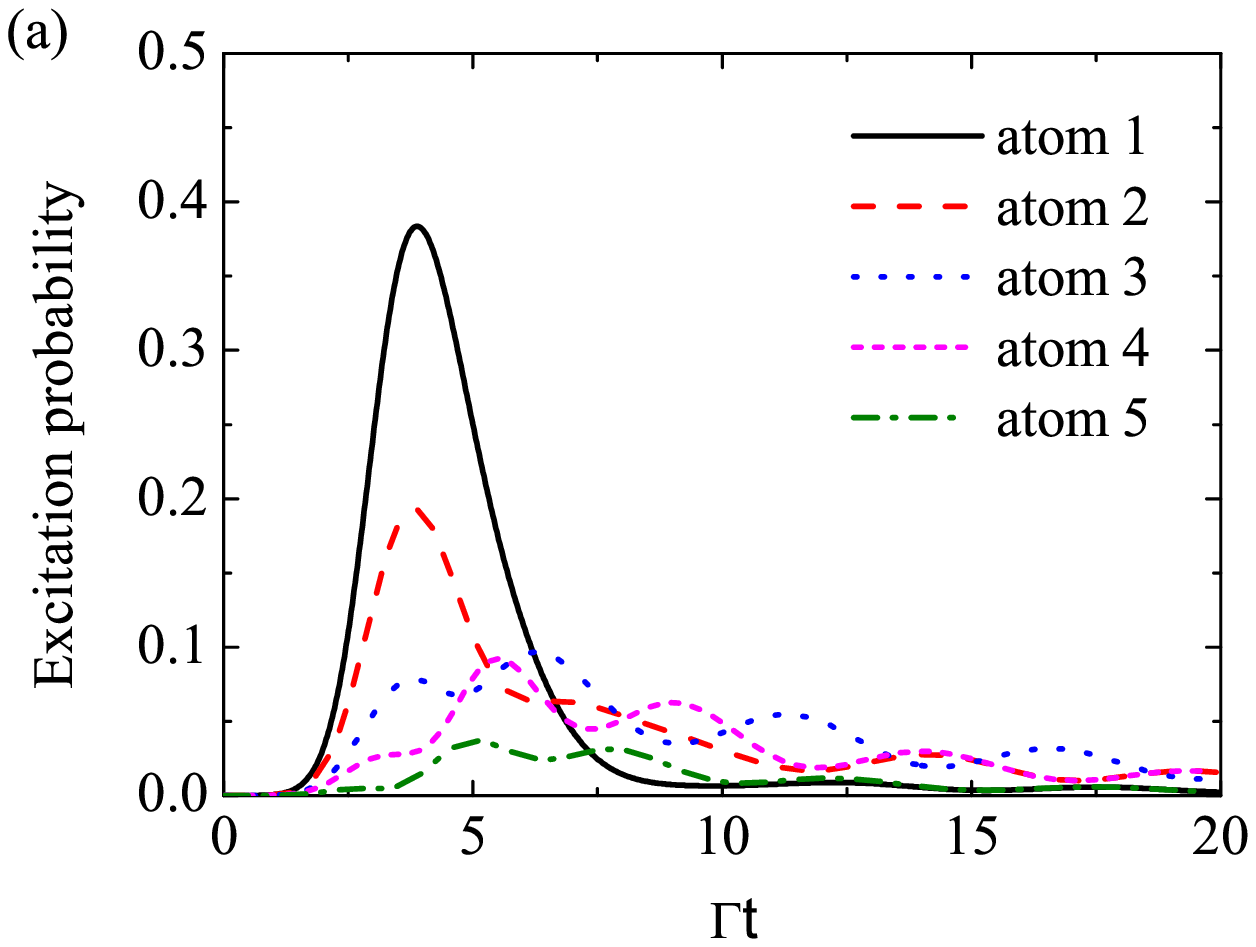}
	\includegraphics[width=0.6\columnwidth]{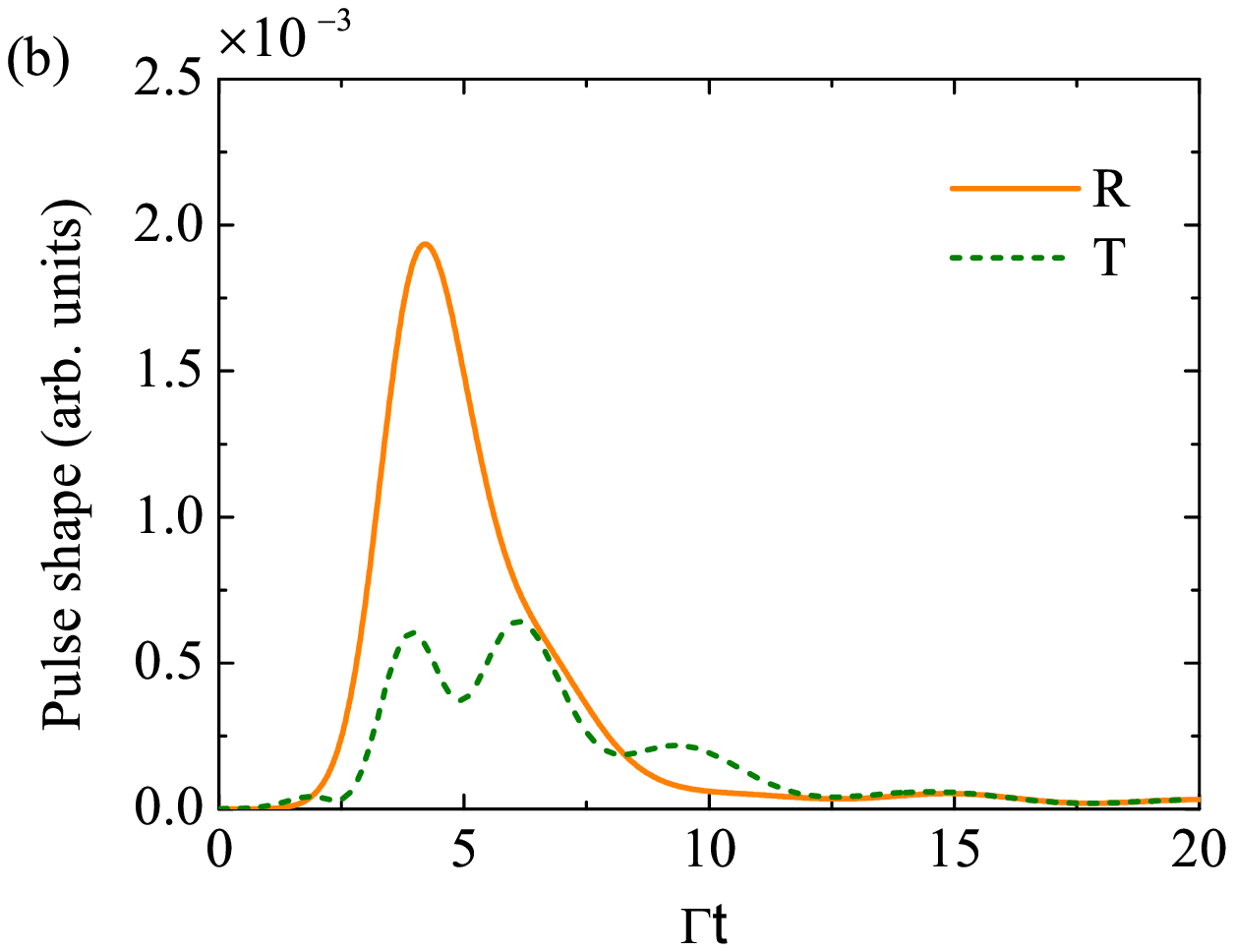}
	\includegraphics[width=0.6\columnwidth]{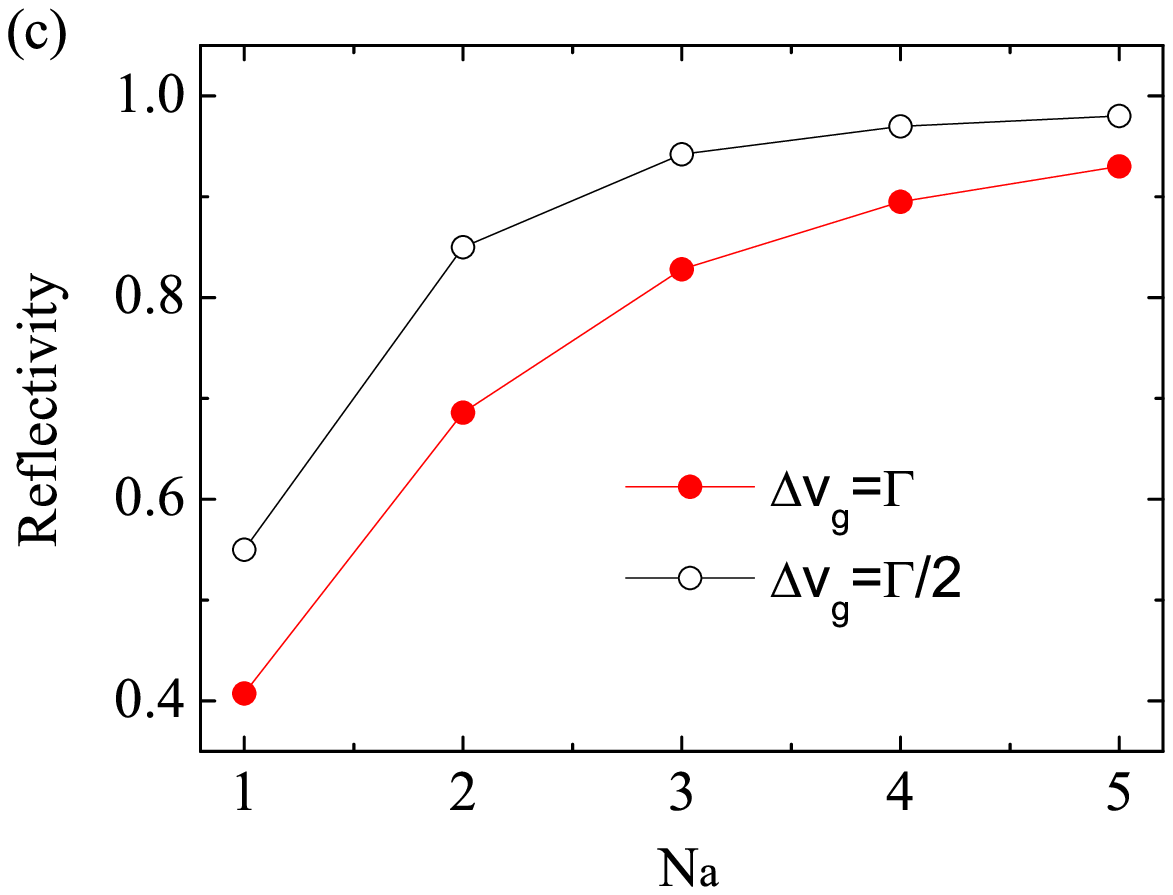}
	\caption{(a) Emitter excitation as a function of time for five emitters. The distance between nearest emitters is $0.25\lambda_{a}$. The pulse spectrum width $\Delta v_{g}=\Gamma$. (b) The reflected and transmitted pulse shapes for the same parameters as (a). (c) The reflectivity as a function of number of emitters for two different incident pulse spectrum widths. The emitter separation is assumed to be $0.5\lambda_{a}$ for (c). }
	\label{7}
\end{figure*}

\subsection{The scattering property at the plane wave limit}

Our theory allows us to study the scattering property of the system at the plane wave limit.
The reflectivity and transmissivity by a single emitter as a function of detuning frequency for the Fock state input and the coherent state input at the plane wave limit are shown as the symbols in Fig. 6(a). It is seen that the reflectivity and transmissivity are  the same for the Fock state input and the coherent state input when the incident photon is a plane wave. When the incident frequency is resonant with the emitter transition frequency, it will be completely reflected back due to  the quantum interference. When the photon frequency is large detuned from the emitter frequency, it can pass through the emitter without being scattering. The widths of the reflectivity and transmissivity depend on the emitter decay rate. The solid lines are the analytical results calculated by the stationary theory \cite{Shen2005} from which we can see that our results here match the previous theoretical results very well. We also find that the reflectivity and transmissivity for a certain frequency is a property of the waveguide-QED system, and it does not depend on the photon statistics of the incident photons.  However, if an incident photon with finite spectrum width, the reflectivity and transmissivity of the pulse can strongly depend on the pulse width and the photon statistics of the incident photons.

In Fig. 6(b), we compare the reflectivity and transmissivity when there is a single emitter or two emitters with separation $a=0.5\lambda_{a}$. The symbols are numerical results calculated by our input-output theory and the solid lines are theoretical results \cite{Liao2015}. We can see that the theoretical results match our numerical results very well which again indicates the validity of our theory. We also find that the reflectivity spectrum when there are two emitters has a broader linewidth than that when there is only a single emitter due to the collective effect.

\subsection{General multiple-emitter case}

In addition to the one or two emitters, our theory can in principle be applied to calculate the dynamics of an arbitrary number of emitters interacting a multi-photon pulse. Here, we take five emitters with nearest neighbor distance $0.25\lambda_{a}$ as an example. The excitation probabilities for the five emitters as a function of time are shown in Fig. 7(a) where we assume that the incident photon pulse is in a coherent state with average photon number 1. We can see that the first emitter has the largest excitation probability, but it is quickly deexcited and can transfer its energy to the other emitters. The other emitters have smaller excitation probabilities, but they can oscillate and last for a period of time much longer than the decay time of single emitter and the incident pulse duration. This is a signature of collective many-body effects where  the collective subradiant states can be generated due to the emitter-emitter interactions and these subradiant states can be populated by the incident photon pulse. The corresponding reflected and transmitted photon pulses are shown in Fig. 7(b). Most energy is reflected and the reflected pulse has a major peak. In contrast, the transmitted pulse has multiple peaks due to quantum interference between the incident field and the reemitted fields by the emitters. The reflectivity as a function of number of emitters is shown in Fig. 7(c) for two different coherent pulse spectrum widths (i.e, $\Delta v_{g}=\Gamma$ and $\Delta v_{g}=\Gamma/2$) where we assume that the emitter distance is $0.5\lambda_{a}$. It is clearly seen that the reflectivity increases when the number of emitter increases and it can approach almost unit when the number of emitter is large. This is another indication of collective effects where superradiant states can be formed. This phenomena can be utilized for atomic mirrors with large frequency bandwidth.

\begin{figure}
	\includegraphics[width=0.49\columnwidth]{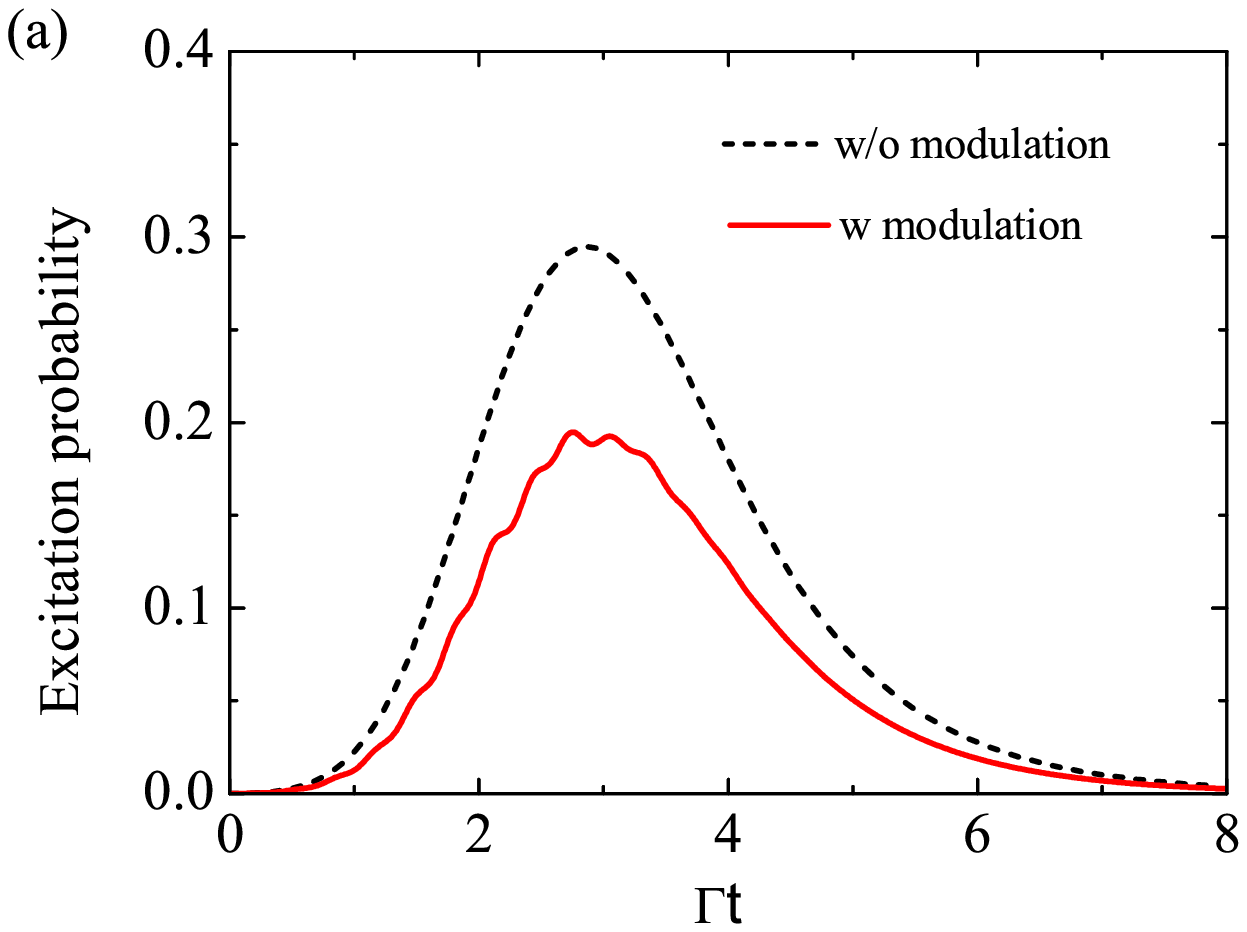}
	\includegraphics[width=0.49\columnwidth]{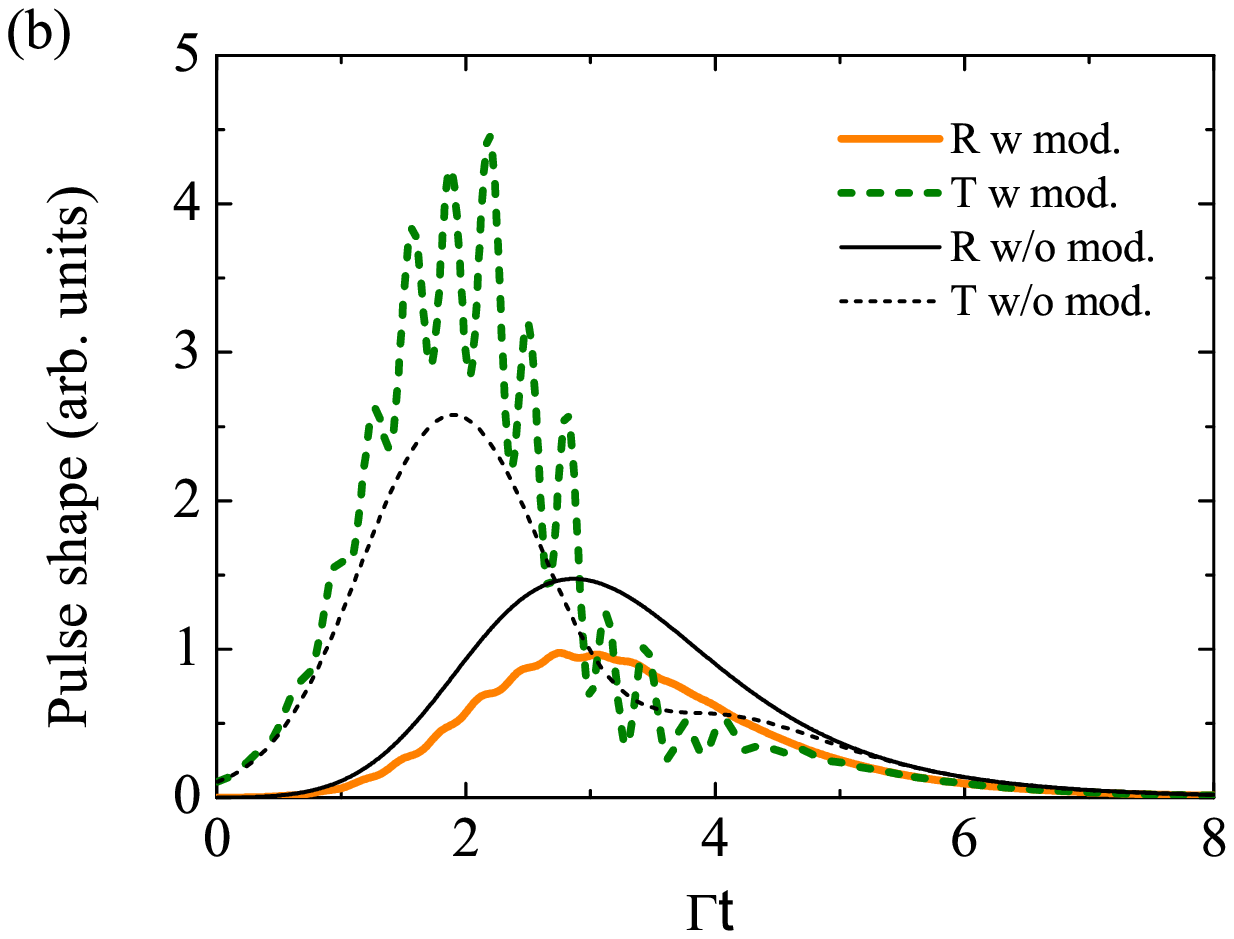}
	\caption{ (a) Emitter excitation as a function of time for a single emitter with frequency modulation (solid red line). For comparison, the excitation without modulation is also shown as the black dashed line. $\varepsilon (t)=10\Gamma \sin (10\Gamma t)$.  The pulse spectrum width $\Delta v_{g}=\Gamma$. (b) The reflected and transmitted pulse shapes for the same parameters as (a).  }
	\label{8}
\end{figure}

\subsection{Emitter frequency modulation}

Our theory also allow us to calculate the transport dynamics when the emitter frequencies are externally modulated. As an example, we consider a single emitter  interacting with a coherent photon pulse. The emitter's frequency is modulated such that $\varepsilon (t)=10\Gamma \sin (10\Gamma t)$. The emitter excitation as a function of time is shown as the red solid line in Fig. 8(a). For comparison, the result without frequency modulation is also plotted as the black dashed line. It is seen that the excitation with modulation is smaller and has some small oscillations.  The corresponding  scattering pulses are shown in Fig. 8(b) where the thicker orange solid line is the reflected pulse and the thicker olive dashed line is the transmitted pulse. We can see that the reflected pulse has small modulations, while the transmitted pulse has very significant modulations. In comparison, the scattering pulses without modulations have smooth shapes (thinner black solid and thinner dashed lines). Hence, by frequency modulation we can realize complicated photon pulse shaping. Since photon pulse shape is very important for high efficient quantum state preparation and transfer \cite{Liao2018}, the theory here can may find important applications in the quantum information.

\section{summary}

In this article we derive master equations for multiphoton interacting with multiple emitters coupled to 1D waveguide. Our theory can be applied to calculate the transport of arbitrary incident photon wavepackets with very general states of light such as coherent state, Fock state and their superpositions. It can also be used to calculate the scattering of  multiple emitters with random distribution and even with external frequency modulation. We compare the dynamics of emitters and scattering pulse shapes when the incident photon pulses are coherent state, single photon state or multiple photon states. With finite incident pulse width, different states of light can induce different system dynamics and different scattering properties. The average reflected photon number by a single emitter decreases when the incident  pulse duration is shorter for both the coherent state input and the Fock state input, but the Fock state input can have higher average reflected photon number than that of the coherent state input with the same average photon number. The results shown here can be useful for single photon generation.   At the plane wave limit,  the reflectivity and transmissivity of the waveguide-QED system for a certain frequency are the same and do not depend on the statistics of the incident photons.  Our theory also allows to study the scattering properties of a photon pulse by emitters with frequency modulations which can be used for photon pulse shaping. We also show that the reflectivity can significantly increase for a broad spectrum due to the collective interaction between the emitters which may be useful for designing atomic mirror with wide bandwidth.  Thus, the theory developed here can become important basics for studying the many-body physics and quantum information applications in the waveguide-QED system. 

\section{ACKNOWLEDGEMENT}
This research is supported by the Key R\&D Program of Guangdong Province (Grant No. 2018B030329001) and startup grants (No. 74130-18841222 and No. 20lgpy163) from Sun Yat-sen University. The research of MSZ is supported by a grant from King Abdulaziz city for Science and Technology (KACST).

\appendix
	
\begin{widetext}	
	
\section{Derivation of Eq. (13) }

On inserting Eqs. (7-10) in the main text into Eq. (2), we can obtain
\begin{align}
\dot{O}_S(t)=&\frac{i}{2}\sum_{j=1}^{N_{a}}\varepsilon_{j}(t)[\sigma_{j}^{z}(t),O_{S}(t)] \nonumber\\
&+i\sum_{j=1}^{N_a}\sum_{k}g_{k}^{j}e^{ikz_{j}}e^{-i\Delta\omega_{k}t} [\sigma_{j}^{+}(t),O_{S}(t)]\Big(a_{k}(0)-i\sum_{l=1}^{N_a}g_{k}^{l*}e^{-ikz_{l}}\int_{0}^{t}\sigma_{l}^{-}(t')e^{i\Delta\omega_{k}t'}dt'\Big )\nonumber \\ &+i\sum_{j=1}^{N_a}\sum_{k}g_{k}^{j*}e^{-ikz_{j}}e^{i\Delta\omega_{k}t}\Big (a_{k}^{\dagger}(0)+i\sum_{l=1}^{N_a}g_{k}^{l}e^{ikz_{l}}\int_{0}^{t}\sigma_{l}^{+}(t')e^{-i\Delta\omega_{k}t'}dt'\Big )[\sigma_{j}^{-}(t),O_{S}(t)] \nonumber \\
&+i\sum_{j=1}^{N_a}\sum_{\vec{q}_{\lambda}}g_{\vec{q}_{\lambda}}^{j}e^{i\vec{q}\cdot \vec{r}_{j}}e^{-i\Delta\omega_{\vec{q}_{\lambda}}t} [\sigma_{j}^{+}(t),O_{S}(t)]\Big(a_{\vec{q}_{\lambda}}(0)-i\sum_{l=1}^{N_a}g_{\vec{q}_{\lambda}}^{l*}e^{-i\vec{q}\cdot \vec{r}_{l}}\int_{0}^{t}\sigma_{l}^{-}(t')e^{i\Delta\omega_{\vec{q}_{\lambda}}t'}dt'\Big )\nonumber \\ &+i\sum_{j=1}^{N_a}\sum_{\vec{q}_{\lambda}}g_{\vec{q}_{\lambda}}^{j*}e^{-i\vec{q}\cdot \vec{r}_{j}}e^{i\Delta\omega_{\vec{q}_{\lambda}}t}\Big (a_{\vec{q}_{\lambda}}^{\dagger}(0)+i\sum_{l=1}^{N_a}g_{\vec{q}_{\lambda}}^{l}e^{i\vec{q}\cdot \vec{r}_{l}}\int_{0}^{t}\sigma_{l}^{+}(t')e^{-i\Delta\omega_{\vec{q}_{\lambda}}t'}dt'\Big )[\sigma_{j}^{-}(t),O_{S}(t)] \nonumber\\
=&\frac{i}{2}\sum_{j=1}^{N_{a}}\varepsilon_{j}(t)[\sigma_{j}^{z}(t),O_{S}(t)]\nonumber\\&+i\sum_{j=1}^{N_a}\sum_{k}g_{k}^{j}e^{ikz_{j}}e^{-i\Delta\omega_{k}t} [\sigma_{j}^{+}(t),O_{S}(t)]a_{k}(0)+i\sum_{j=1}^{N_a}\sum_{k}g_{k}^{j*}e^{-ikz_{j}}e^{i\Delta\omega_{k}t}a_{k}^{\dagger}(0)[\sigma_{j}^{-}(t),O_{S}(t)]  \nonumber\\
&+i\sum_{j=1}^{N_a}\sum_{\vec{q}_{\lambda}}g_{\vec{q}_{\lambda}}^{j}e^{i\vec{q}\cdot \vec{r}_{j}}e^{-i\Delta\omega_{\vec{q}_{\lambda}}t} [\sigma_{j}^{+}(t),O_{S}(t)]a_{\vec{q}_{\lambda}}(0)+i\sum_{j=1}^{N_a}\sum_{\vec{q}_{\lambda}}g_{\vec{q}_{\lambda}}^{j*}e^{-i\vec{q}\cdot \vec{r}_{j}}e^{i\Delta\omega_{\vec{q}_{\lambda}}t}a_{\vec{q}_{\lambda}}^{\dagger}(0)[\sigma_{j}^{-}(t),O_{S}(t)]  \nonumber\\
&+\sum_{jl}^{N_a}\sum_{k}g_{k}^{j}g_{k}^{l*}e^{ik(z_{j}-z_{l})}e^{-i\Delta\omega_{k}t} [\sigma_{j}^{+}(t),O_{S}(t)]\int_{0}^{t}\sigma_{l}^{-}(t')e^{i\Delta\omega_{k}t'}dt' \nonumber \\ 
&-\sum_{jl}^{N_a}\sum_{k}g_{k}^{j}g_{k}^{l*}e^{-ik(z_{j}-z_{l})}e^{i\Delta\omega_{k}^{j}(t)t}\int_{0}^{t}\sigma_{l}^{+}(t')e^{-i\Delta\omega_{k}t'}dt'[\sigma_{j}^{-}(t),O_{S}(t)] \nonumber \\
&+\sum_{jl}^{N_a}\sum_{\vec{q}_{\lambda}}g_{\vec{q}_{\lambda}}^{j}g_{\vec{q}_{\lambda}}^{l*}e^{i\vec{q}\cdot (\vec{r}_{j}-\vec{r}_{l})}e^{-i\Delta\omega_{\vec{q}_{\lambda}}t} [\sigma_{j}^{+}(t),O_{S}(t)]\int_{0}^{t}\sigma_{l}^{-}(t')e^{i\Delta\omega_{\vec{q}_{\lambda}}t'}dt' \nonumber \\ 
&-\sum_{jl}^{N_a}\sum_{\vec{q}_{\lambda}}g_{\vec{q}_{\lambda}}^{j*}g_{\vec{q}_{\lambda}}^{l}e^{-i\vec{q}\cdot (\vec{r}_{j}-\vec{r}_{l})}e^{i\Delta\omega_{\vec{q}_{\lambda}}t} \int_{0}^{t}\sigma_{l}^{+}(t')e^{-i\Delta\omega_{\vec{q}_{\lambda}}t'}dt'[\sigma_{j}^{-}(t),O_{S}(t)].
\end{align}
According to the Weisskopf-Wigner approximation, we have \cite{Liao2016}
\begin{align}
\sum_{k}g_{k}^{j}g_{k}^{l*}e^{ik(z_{j}-z_{l})}e^{-i\Delta\omega_{k}t}e^{i\Delta\omega_{k}t'} &=\frac{\sqrt{\Gamma_{j}\Gamma_{l}}}{2}e^{ik_a |z_{jl}|}e^{i\varepsilon_{j}(t)t-i\varepsilon_{l}(t')t'}\delta[t'-(t-\frac{|z_{jl}|}{v_g})], \\
\sum_{k}g_{k}^{j}g_{k}^{l*}e^{-ik(z_{j}-z_{l})}e^{i\Delta\omega_{k}t}e^{-i\Delta\omega_{k}t'} &=\frac{\sqrt{\Gamma_{j}\Gamma_{l}}}{2}e^{-ik_a |z_{jl}|}e^{-i\varepsilon_{j}(t)t+i\varepsilon_{l}(t')t'}\delta[t'-(t-\frac{|z_{jl}|}{v_g})], \\
\sum_{\vec{q}_{\lambda}}g_{\vec{q}_{\lambda}}^{j}g_{\vec{q}_{\lambda}}^{l*}e^{i\vec{q}\cdot (\vec{r}_{j}-\vec{r}_{l})}e^{-i\Delta\omega_{\vec{q}_{\lambda}}t}e^{i\Delta\omega_{\vec{q}_{\lambda}}t'}&=\Omega_{jl}e^{i\varepsilon_{j}(t)t-i\varepsilon_{l}(t')t'}\delta [t'-(t-\frac{|r_{jl}|}{v_{g}})], \\
\sum_{\vec{q}_{\lambda}}g_{\vec{q}_{\lambda}}^{j*}g_{\vec{q}_{\lambda}}^{l}e^{-i\vec{q}\cdot (\vec{r}_{j}-\vec{r}_{l})}e^{i\Delta\omega_{\vec{q}_{\lambda}}t}e^{-i\Delta\omega_{\vec{q}_{\lambda}}t'}&=\Omega_{jl}^{*}e^{-i\varepsilon_{j}(t)t+i\varepsilon_{l}(t')t'}\delta [t'-(t-\frac{|r_{jl}|}{v_{g}})], 
\end{align}
where $\Gamma_{i}=\frac{2L}{v_{g}}|g_{k_{0}}^{i}|^{2}$ with   $g_{k_{0}}^{i}=\sqrt{\frac{\Gamma_{i} v_{g}}{2L}}$ and $\Omega_{jl}=\frac{3\sqrt{\gamma_{j}\gamma_{l}}}{4}[\sin^{2}\phi\frac{-i}{k_{a}r_{jl}}+(1-3\cos^{2}\phi)(\frac{1}{(k_{a}r_{jl})^2}+\frac{i}{(k_{a}r_{jl})^{2}})]e^{ik_{a}r_{jl}}$ with $r_{jl}=|\vec{r_{j}}-\vec{r_{l}}|$. 

To proceed, we assume that the emitters are close such that $z_{ij}/v_g\ll 1/\Gamma$, we can approximate that $\sigma_{j}^{-}(t-\frac{z_{jl}}{v_{g}})\approx \sigma_{j}^{-}(t)$ in the rotating frame. Indeed, this is the usual case. For example, if $v_g\sim 10^{8}m/s$ and $\Gamma\sim 10^{8}Hz$, we require that the distance between the emitters $z_{ij}\ll 1m$ which is the usual case. By doing this approximation, Eq. (13) then becomes
\begin{align}
\dot{O}_S(t)=&\frac{i}{2}\sum_{j=1}^{N_{a}}\varepsilon_{j}[\sigma_{j}^{z}(t),O_{s}(t)]+ i\sum_{j=1}^{N_a} \sqrt{\frac{\Gamma_{j}}{2}}[\sigma_{j}^{+}(t),O_{S}(t)][a_{j}(t)+b_{j}(t)]
+i\sum_{j=1}^{N_a}\sqrt{\frac{\Gamma_{j}}{2}}[a_{j}^{\dagger}(t)+b_{j}^{\dagger}(t)][\sigma_{j}^{-}(t),O_{S}(t)]  \nonumber \\ 
&+\sum_{jl}\Lambda_{jl}[\sigma_{j}^{+}(t),O_{S}(t)]\sigma_{l}^{-}(t)
-\sum_{jl}\Lambda_{jl}^{*}\sigma_{l}^{+}(t)[\sigma_{j}^{-}(t),O_{S}(t)],
\end{align}
where $a_{j}(t)=\sqrt{\frac{v_{g}}{2\pi}}\int_{0}^{\infty}e^{ikz_{j}}a_{k}(0)e^{-i\delta\omega_{k}t}dk$ 
is the absorption of the incident waveguide photons and $b_{j}(t)=\sqrt{\frac{v_{g}}{2\pi}}\int \int\int e^{i\vec{q}_{\lambda}\cdot \vec{r}_{j}}a_{\vec{q}_{\lambda}}(0)e^{-i\delta\omega_{\vec{q}_{\lambda}}t}d^{3}\vec{q}_{\lambda}$ is the absorption of the incident nonguided photons. The collective interaction between the emitters is given by \cite{Liao2016}
\begin{equation}
\Lambda_{jl}=\frac{\sqrt{\Gamma_{j}\Gamma_{l}}}{2}e^{ik_{a}|z_{jl}|}+\frac{3\sqrt{\gamma_{j}\gamma_{l}}}{4}[\sin^{2}\phi\frac{-i}{k_{a}r_{jl}}+(1-3\cos^{2}\phi)(\frac{1}{(k_{a}r_{jl})^2}+\frac{i}{(k_{a}r_{jl})^{2}})]e^{ik_{a}|r_{jl}|}. 
\end{equation}

From Eq. (A6), we can derive a corresponding master equation for the emitters. Since $Tr_{S+R}[O_S(t)\rho]=Tr_S[O_S\rho_S(t)]$ where $\rho_S(t)=Tr_{R}[\rho(t)]$, we have
\begin{align}
&Tr_S[O_S\dot{\rho_S}(t)]\nonumber \\
&=Tr_{S+R}[\dot{O}_S(t)\rho] \nonumber \\
&=\frac{i}{2}\sum_{j=1}^{N_{a}}\varepsilon_{j}(t)Tr_{S+R}\{[\sigma_{j}^{z}(t),O_{S}(t)]\rho\}\nonumber\\&+i\sum_{j=1}^{N_a}\sqrt{\frac{\Gamma_{j}}{2}}Tr_{S+R}\{ [\sigma_{j}^{+}(t),O_{S}(t)][a_{j}(t)+b_{j}(t)] \rho\}
+i\sum_{j=1}^{N_a}\sqrt{\frac{\Gamma_{j}}{2}}Tr_{S+R}\{[a_{j}^{\dagger}(t)+b_{j}^{\dagger}(t)] [\sigma_{j}^{-}(t),O_{S}(t)]\rho\} \nonumber\\ 
&+\sum_{jl}\Lambda_{jl}Tr_{S+R}\{ [\sigma_{j}^{+}(t),O_{S}(t)]\sigma_{l}^{-}(t)\rho\}
-\sum_{jl}\Lambda^{*}_{jl} Tr_{S+R}\{ \sigma_{l}^{+}(t)[\sigma_{j}^{-}(t),O_{S}(t)]\rho\} \nonumber \\
&=-\frac{i}{2}\sum_{j=1}^{N_{a}}\varepsilon_{j}(t)Tr_{S}\{O_{S}[\sigma_{j}^{z},\rho_{S}(t)]\} + i\sum_{j=1}^{N_a}\sqrt{\frac{\Gamma_{j}}{2}}Tr_{S}\{ O_{S}[\rho_{in}^{j}(t),\sigma_{j}^{+}]\}
+i\sum_{j=1}^{N_a}\sqrt{\frac{\Gamma_{j}}{2}}Tr_{S}\{ O_{S} [\rho_{in}^{j\dagger}(t), \sigma_{j}^{-}]\} \nonumber \\
&-\sum_{jl}Tr_{S}\{ O_{S} [\sigma_{j}^{+}\sigma_{l}^{-}\rho_{S}(t)-\sigma_{l}^{-}\rho_{S}(t)\sigma_{j}^{+}]\}-\sum_{jl}\Lambda_{jl}^{*}Tr_{S}\{ O_{S} [\rho_{S}(t)\sigma_{l}^{+}\sigma_{j}^{-}-\sigma_{j}^{-}\rho_{S}(t)\sigma_{l}^{+}]\},
\end{align}
where $\rho_{in}^{j}(t)=Tr_{R}\{U(t)[a_{j}(t)+b_{j}(t)]\rho(0) U^{\dagger}(t)\}$ is the contribution from the incident sources. In this paper, we consider that the incident photon is coming from the waveguide photons and the non-guided modes are initially in the vacuum. Since $a_{\vec{q}_{\lambda}}(0)|0\rangle =0$, we have $b_{j}(t)\rho(0)=0$ and therefore $\rho_{in}^{j}(t)=Tr_{R}\{U(t)a_{j}(t)\rho(0) U^{\dagger}(t)\}$ is due to the contribution of the incident waveguide photons. Comparing both size of Eq. (A8), we can obtain the master equation for the system density matrix given by 
\begin{align}
\dot{\rho}_S(t)=& -\frac{i}{2}\sum_{j=1}^{N_{a}}\varepsilon_{j}(t)[\sigma_{j}^{z},\rho_{S}(t)]-i\sum_{j=1}^{N_a}\sqrt{\frac{\Gamma_{j}}{2}} [\sigma_{j}^{+},\rho_{in}^{j}(t)]
-i\sum_{j=1}^{N_a}\sqrt{\frac{\Gamma_{j}}{2}}[\sigma_{j}^{-},\rho_{in}^{j\dagger}(t)] \nonumber \\ &-i\sum_{jl}\text{Im}(\Lambda_{jl})[\sigma_j^{+}\sigma_l^{-},\rho_{S}(t)] 
-\sum_{jl}\text{Re}(\Lambda_{jl})[\sigma_{j}^{+}\sigma_{l}^{-}\rho_{S}(t)+\rho_{S}(t)\sigma_{j}^{+}\sigma_{l}^{-}-2\sigma_{l}^{-}\rho_{S}(t)\sigma_{j}^{+}],  
\end{align}
which is the master equation shown in Eq. (13) in the main text.

\section{Derivation of the input-output relations}

From Eqs. (15-19) in the main text, we can obtain
\begin{align}
a_{out}^{R}(t)&=a_{in}^{R}(t-z_{N}/v_{g})-i\sum_{j=1}^{N_a}\sqrt{\frac{\Gamma_{j}v_{g}}{4\pi}}\int_{0}^{t_f}\sigma_{j}^{-}(t')dt'\sqrt{\frac{v_{g}}{2\pi}}\int_{0}^{\infty}e^{i\delta kz_{N}}e^{-ikz_{j}}e^{i\Delta\omega_{k}(t'-t)}dk \nonumber \\
&=a_{in}^{R}(t-z_{N}/v_{g})-i\sum_{j=1}^{N_a}\sqrt{\frac{\Gamma_{j}v_{g}}{4\pi}}\sqrt{\frac{v_{g}}{2\pi}}e^{-i k_{a}z_{j}}\int_{0}^{t_f}\sigma_{j}^{-}(t')dt'\int_{0}^{\infty}e^{i\delta k(z_{N}-z_{j})}e^{i\delta k v_{g}(t'-t)}dk  \nonumber \\
&=a_{in}^{R}(t-z_{N}/v_{g})-i\sum_{j=1}^{N_a}\sqrt{\frac{\Gamma_{j}v_{g}}{4\pi}}\sqrt{\frac{v_{g}}{2\pi}}e^{-i k_{a}z_{j}}\int_{0}^{t_f}\sigma_{j}^{-}(t')dt'\int_{-k_{0}}^{\infty}e^{i\delta kz_{Nj}}e^{i\delta k v_{g}(t'-t)}d\delta k  \nonumber \\
&=a_{in}^{R}(t-z_{N}/v_{g})-i\sum_{j=1}^{N_a}\sqrt{\frac{\Gamma_{j}v_{g}}{4\pi}}\sqrt{\frac{v_{g}}{2\pi}}e^{-i k_{a}z_{j}}\int_{0}^{t_f}\sigma_{j}^{-}(t')dt'\int_{-\infty}^{\infty}e^{i\delta kz_{Nj}}e^{i\delta k v_{g}(t'-t)}d\delta k  \nonumber \\
&=a_{in}^{R}(t-z_{N}/v_{g})-i\sum_{j=1}^{N_a}\sqrt{\frac{\Gamma_{j}v_{g}}{4\pi}}\sqrt{\frac{v_{g}}{2\pi}}e^{-i k_{a}z_{j}}\int_{0}^{t_f}\sigma_{j}^{-}(t')\frac{2\pi}{v_{g}}\delta(t'-t+z_{Nj}/v_{g})dt'\nonumber \\
&=a_{in}^{R}(t-z_{N}/v_{g})-i\sum_{j=1}^{N_a}\sqrt{\frac{\Gamma_{j}}{2}}e^{-i k_{a}z_{j}}\sigma_{j}^{-}(t-z_{Nj}/v_{g})\nonumber \\ &\approx a_{in}^{R}(t-z_{N}/v_{g})-i\sum_{j=1}^{N_a}\sqrt{\frac{\Gamma_{j}}{2}}e^{-i k_{a}z_{j}}\sigma_{j}^{-}(t),
\end{align}
where $z_{Nj}=z_{N}-z_{j}$.
Similarly, we have
\begin{align}
a_{out}^{L}(t)&=a_{in}^{L}(t+z_{1}/v_{g})-i\sum_{j=1}^{N_a}\sqrt{\frac{\Gamma_{j}v_{g}}{4\pi}}\int_{0}^{t_f}\sigma_{j}^{-}(t')dt'\sqrt{\frac{v_{g}}{2\pi}}\int_{-\infty}^{0}e^{-i\delta kz_{1}}e^{-ikz_{j}}e^{i\Delta\omega_{k}(t'-t)}dk \nonumber \\
&=a_{in}^{L}(t+z_{1}/v_{g})-i\sum_{j=1}^{N_a}\sqrt{\frac{\Gamma_{j}v_{g}}{4\pi}}\sqrt{\frac{v_{g}}{2\pi}}e^{i k_{a}z_{j}}\int_{0}^{t_f}\sigma_{j}^{-}(t')dt'\int_{0}^{\infty}e^{i\delta kz_{j1}}e^{i\delta k v_{g}(t'-t)}d(-k)  \nonumber \\
&=a_{in}^{L}(t+z_{1}/v_{g})-i\sum_{j=1}^{N_a}\sqrt{\frac{\Gamma_{j}v_{g}}{4\pi}}\sqrt{\frac{v_{g}}{2\pi}}e^{i k_{a}z_{j}}\int_{0}^{t_f}\sigma_{j}^{-}(t')dt'\int_{-k_{0}}^{\infty}e^{i\delta kz_{j1}}e^{i\delta k v_{g}(t'-t)}d\delta k  \nonumber \\
&=a_{in}^{L}(t+z_{1}/v_{g})-i\sum_{j=1}^{N_a}\sqrt{\frac{\Gamma_{j}v_{g}}{4\pi}}\sqrt{\frac{v_{g}}{2\pi}}e^{i k_{a}z_{j}}\int_{0}^{t_f}\sigma_{j}^{-}(t')dt'\int_{-\infty}^{\infty}e^{i\delta kz_{j1}}e^{i\delta k v_{g}(t'-t)}d\delta k  \nonumber \\
&=a_{in}^{L}(t+z_{1}/v_{g})-i\sum_{j=1}^{N_a}\sqrt{\frac{\Gamma_{j}v_{g}}{4\pi}}\sqrt{\frac{v_{g}}{2\pi}}e^{i k_{a}z_{j}}\int_{0}^{t_f}\sigma_{j}^{-}(t')\frac{2\pi}{v_{g}}\delta(t'-t+z_{j1}/v_{g})dt'\nonumber \\
&=a_{in}^{L}(t+z_{1}/v_{g})-i\sum_{j=1}^{N_a}\sqrt{\frac{\Gamma_{j}}{2}}e^{i k_{a}z_{j}}\sigma_{j}^{-}(t-z_{j1}/v_{g})  \nonumber \\
&\approx a_{in}^{L}(t+z_{1}/v_{g})-i\sum_{j=1}^{N_a}\sqrt{\frac{\Gamma_{j}}{2}}e^{i k_{a}z_{j}}\sigma_{j}^{-}(t). 
\end{align}
Eqs. (B1) and (B2) are the input-output relations of the system from which we can calculate the field scattering properties of this system.

\section{Derivation of Eq. (35)}

We can derive a dynamical equation for $\rho_{01}(t)$ using similar method as deriving $\rho_{S}(t)$. Since $Tr_{S+R}[O_S\rho_{01}(t)]=Tr_{S+R}[O_S(t)\rho_{01}(0)]$ where $\rho_{01}(0)=\rho_{S}(0) \otimes |0\rangle \langle \Psi_F|$, we have
\begin{align}
&Tr_{S+R}[O_S\dot{\rho}_{01}(t)] \nonumber\\
=&Tr_{S+R}[\dot{O}_S(t)\rho_{01}(0)] \nonumber \\
=&\frac{i}{2}\sum_{j=1}^{N_{a}}\varepsilon_{j}(t)Tr_{S+R}\{[\sigma_{j}^{z}(t),O_{s}(t)]\rho_{01}(0)\} \nonumber \\
&+i\sum_{j=1}^{N_a}\sqrt{\frac{\Gamma_{j}}{2}}Tr_{S+R}\{ [\sigma_{j}^{+}(t),O_{S}(t)]a_{in}^{j}(t-z_{j}/v_{g}) \rho_{01}(0)\}
+i\sum_{j=1}^{N_a}\sqrt{\frac{\Gamma_{j}}{2}}Tr_{S+R}\{ a_{in}^{j\dagger}(t-z_{j}/v_{g})[\sigma_{j}^{-}(t),O_{S}(t)]\rho_{01}(0)\} \nonumber\\ 
&+\sum_{jl}\Lambda_{jl}Tr_{S+R}\{ [\sigma_{j}^{+}(t),O_{S}(t)]\sigma_{l}^{-}(t)\rho\}
-\sum_{jl}\Lambda^{*}_{jl}Tr_{S+R}\{ \sigma_{l}^{+}(t)[\sigma_{j}^{-}(t),O_{S}(t)]\rho\} \nonumber \\
=&-\frac{i}{2}\sum_{j=1}^{N_{a}}\varepsilon_{j}(t)Tr_{S+R}\{[\sigma_{j}^{z},\rho_{01}(t)]O_{s}\}
-i\sum_{j=1}^{N_a}\sqrt{\frac{\Gamma_{j}}{2}}\alpha_{j}^{*}(t-z_{j}/v_{g})Tr_{S+R}\{ O_{S} [\sigma_{j}^{-},\rho_{00}(t)]\} \nonumber \\
&-\sum_{jl}\Lambda_{jl}Tr_{S+R}\{ O_{S} [\sigma_{j}^{+}\sigma_{l}^{-}\rho_{01}(t)-\sigma_{l}^{-}\rho_{01}(t)\sigma_{j}^{+}]\}
-\sum_{jl}\Lambda_{jl}^{*}Tr_{S+R}\{ O_{S} [\rho_{01}(t)\sigma_{l}^{+}\sigma_{j}^{-}-\sigma_{j}^{-}\rho_{01}(t)\sigma_{l}^{+}]\}, 
\end{align}
where $\rho_{00}(t)=U(t)\rho_S \otimes |0\rangle \langle 0|U^{\dagger}(t)$. Comparing both sides, we have 
\begin{equation}
\dot{\rho}_{01}^{S}(t)=-\frac{i}{2}\sum_{j=1}^{N_{a}}\varepsilon_{j}(t)[\sigma_{j}^{z},\rho_{01}^{S}(t)]  -i\sum_{j=1}^{N_a}\sqrt{\frac{\Gamma_{j}v_{g}}{2L}}\alpha_{j}^{*}(t) [\sigma_{j}^{-},\rho_{00}^{S}(t)] 
-i\sum_{jl}\text{Im}(\Lambda_{jl}) [\sigma_{j}^{+}\sigma_{l}^{-},\rho_{01}^{S}(t)] -\mathcal{L}[\rho_{01}^{S}(t)],
\end{equation}
where $\rho_{00}^{S}(t)=Tr_{R}[\rho_{00}(t)]$.

\end{widetext}

\end{document}